\documentclass{article}
\usepackage{jcappub}
\usepackage{tensor}
\usepackage[normalem]{ulem}

\def\coordDisp{\Delta}
\def\bare{{\rm bare}}
\def\newt{{\rm (newt)}}
\def\longitudinal{{\rm (long)}}
\def\synchronous{{\rm (syn)}}
\def\comoving{{\rm (comov)}}
\def\l{\left}
\def\r{\right}

\def\hub{{\mathcal H}}

\title{Initial conditions for cosmological N-body simulations of the scalar sector of theories of Newtonian, Relativistic and Modified Gravity}
\author{Wessel Valkenburg,}
\author{Bin  Hu}
\affiliation{Instituut-Lorentz, Universiteit Leiden, P.O. Box 9506, 2300 RA Leiden, The Netherlands}

\emailAdd{valkenburg@lorentz.leidenuniv.nl}
\emailAdd{hu@lorentz.leidenuniv.nl}

\abstract{
{  We present a 
description for setting initial particle displacements and field values for simulations of arbitrary metric theories of gravity, for perfect and imperfect fluids with arbitrary characteristics. We extend the Zel'dovich Approximation to nontrivial theories of gravity, and show how scale dependence implies curved particle paths, even in the entirely linear regime of perturbations. For a viable choice of Effective Field Theory of Modified Gravity, initial conditions set at high redshifts are affected at the level of up to 5\% at Mpc scales, which exemplifies the importance of going beyond $\Lambda$-Cold Dark Matter initial conditions for modifications of gravity outside of the quasi-static approximation. In addition, we show initial conditions for a simulation where a scalar modification of gravity is modelled in a Lagrangian particle-like description. Our description paves the way for simulations and mock galaxy catalogs under theories of gravity beyond the standard model, crucial for progress towards precision tests of gravity and cosmology.}
}

\begin{document}
\maketitle

\section{Introduction}

The expansion history of the universe seems to be a smooth function of time, which can hence only provide a handle on a limited number of parameters in the cosmological model. In order to further quantify Inflation, Dark Matter and Dark Energy, and distinguish a variety of candidate models, hope is placed on probing the nonlinear growth of perturbations in the universe, considered the ongoing preparations for the Euclid satellite~\cite{2014SPIE.9143E..0HL}.

The current state of the art for nonlinear structure formation in the universe is the numerical simulation of gravitational clustering of a large fixed number of particles in a finite sized box: N-body simulations. A plethora of codes solves Newtonian dynamics endowed with a friction term that accounts for the cosmic expansion ({\em e.g.}~\cite{Teyssier:2001cp, Springel:2005mi,Bryan:2013hfa}). Progress is made in relativistic simulations as well~\cite{Adamek:2014xba}. The next layer of complexity is the addition of perturbations in the exotic form of matter mentioned above, Dark Energy, or often under the alternative moniker of Modified Gravity~\cite{Llinares:2008ce, Baldi:2008ay, Zhao:2010qy, Baldi:2013iza, Llinares:2013jza}.

{  
	In order to push simulations to the next level, accounting for dynamical modifications of gravity or Dark Energy at arbitrary redshifts, an obvious first step is to discuss the starting point of such simulations, which is the goal of this paper. 
}

Numerical simulations are inevitably discrete in space and time, and it is not clear what the consequences of this discreteness are for the results, or how closely the results reflect nonlinear structure in a continuous universe~\cite{Pen:1997up,Baertschiger:2001eu,Joyce:2004qz,Sirko:2005uz,Prunet:2008fv,Joyce:2008kg,Carron:2014wea,Colombi:2014zga}. Such discreteness effects are not the subject of this paper. {  However, in Appendix~\ref{sec:discrete}, we summarise the known methods for realising discrete samples of continuous fields, and we list the choices that the simulator needs to make.}

In most of the literature on simulations, priority is given to the details and technicalities of solving for the nonlinear dynamics. Setting the initial conditions is always done using Zel'dovich' approximation (explained in the main text), sometimes up to second order~\cite{Crocce:2006ve}. Most importantly, to our knowledge, {\em all} simulations set their initial conditions at a time when no perturbations in exotic matter are present. However, if one wants a more complete description, the full parameter range of the exotic models needs to be tested, including the range in which Dark Energy has significant dynamical perturbations (see~\cite{Clifton:2011jh} for a review); where the quasi-static approximation for its perturbations breaks down~\cite{Noller:2013wca,Brito:2014ifa,Sawicki:2015zya,Winther:2015pta}.

The aim of this paper is to describe all that is necessary to set the correct initial conditions for arbitrary cosmological simulations under arbitrary metric theories of gravity, varying from the inclusion of only dust to the inclusion of species that in their linear description are fully imperfect fluids. {   For the sake of clarity, we include discussions of previously known topics where necessary, providing the relevant references.} We do not address the subject of setting initial conditions for matter species that do not simply form a sheet in 6-dimensional phase space, such as for example standard-model neutrinos and photons above a certain resolution, which are described accurately by linear perturbations although not by fluid dynamics. {  We release a code, {\sc FalconIC}~\footnote{{\url{http://falconb.org}}}, which can be used to generate discrete initial conditions of any cosmological {\em fluid}. The code links against any version of both Boltzmann codes {\sc camb}~\cite{Lewis:1999bs} and {\sc class}~\cite{Blas:2011rf}, including {\sc EFTCamb}~\cite{Hu:2013twa,Raveri:2014cka}, such that no separate running of those codes is necessary, and initial conditions for any fluid can be generated at arbitrary scales, of arbitrary size (fully parallelised using MPI and OpenMP), for arbitrary cosmological parameters.  }

Simulations discretise the cosmological fluids in a regime where the fluid description is still valid, on the onset of shell crossing and the onset of the need for a particle description, when the sheet in phase space starts folding or even tearing. Therefore, at the first time steps, N-body simulations should reproduce the linear theory described by fluid dynamics.
{  However, the fact that the linear discussion in this paper addresses (imperfect) fluids, does not at all imply that these discrete realisations can only be used for fluids. In other words, the Effective Field Theory of Modified Gravity describes some matter which in the linear regime may look like an imperfect fluid. Particles whose free-streaming length is smaller than the simulation resolution, but whose dynamics is described as in imperfect fluid, such as standard model neutrinos, can be realised just as well following our description.}

In summary, {  the following points are the ingredients for discrete realisations of arbitrary imperfect fluids},
\begin{itemize}

	\item Any linearly perturbed quantity (`charge') can be translated into a displacement field of equal charge vertices, by a coordinate transformation whose Jacobian equals the original charge perturbations.

	\item The scalar quantity that defines the positions of particles with velocity $u^\mu$, for an arbitrary fluid, is $n^{\bare}\equiv-\int d\tau\, \nabla_i u^i$, where $i$ runs over spatial coordinates.

	\item For newtonian simulations, positions are given by $n\equiv  n^\bare / \sqrt{g^{(3)}}$, with $g^{(3)}$ the determinant of the spatial part of the metric.

	\item In the presence of pressure or heat flux, the particles have varying masses $m$ and internal energies $T$, defined by
	\begin{align}
	m =& \frac{\bar \rho}{\bar n}\left( 1 + \Delta_\rho - \Delta_n \right) =  \frac{\bar \rho}{\bar n}\left( 1 + \Delta_\rho^{\bare} - \Delta_n^{\bare} \right), \\
	T =& \frac{\bar P}{\bar n}\left( 1 + \Delta_P - \Delta_n \right) =  \frac{\bar P}{\bar n}\left( 1 + \Delta_P^{\bare} - \Delta_{n}^{\bare} \right),
\end{align}
with $\Delta_a\equiv \delta a / \bar a$ for any quantity $a$ with background value $\bar a$.

	\item For adiabatic perturbations, all quantities are set by the same random seed multiplied by their respective transfer functions. Isocurvature perturbations are introduced by combining multiple random seeds. 

	\item The linear displacement field for a dust fluid in a universe endowed with a scalar modification of gravity is scale dependent, and hence the trajectories of particles are not straight lines.
	
	\item The scalar modification itself can be described in both Lagrangian and Eulerian representations, whichever of the two may prove more convenient in nonlinear simulations.
\end{itemize}

{  The bending of Dark Matter trajectories at the linear level, implies that simulation codes such as Ramses~\cite{Teyssier:2001cp} or Enzo~\cite{Bryan:2013hfa} which only take particle velocities as input, need to be adapted to include particle positions as well, since particle positions are no longer simply a time-dependent function times their velocities.}

In section~\ref{sec:anyPlainDisp} we start by explaining how an arbitrary density field is related to a displacement field, under the assumption of absence of vorticity\footnote{One can add vorticity by means of several scalar fields~\cite{Schutz:1970my}.}. In section~\ref{sec:mainIC} we then show how to define the displacement field under an arbitrary metric for an arbitrary (im)perfect fluid, with the correct velocities emerging. In section~\ref{sec:nrlim}, we briefly comment on the Newtonian approximation, necessary for Newtonian simulations. In section~\ref{sec:qsa} we show how particle trajectories of linear perturbations do not follow straight lines, already in the quasi-static approximation. Finally, in section~\ref{sec:demgeftLPT} we apply our prescription to a parameterisation of the Effective Field Theory of Modified Gravity, in synchronous comoving coordinates.

Throughout this paper, we will refer to the Conformal Newtonian gauge as the Longitudinal gauge, since we prefer to reserve the word Newtonian for Newtonian N-body simulations.
 We use units with the speed of light $c=1$, and we use the Einstein summation convention.
All equations with perturbative quantities are expanded up to linear order.

\section{Density--displacement duality\label{sec:anyPlainDisp}}

\begin{figure}
\includegraphics[width=0.24\textwidth]{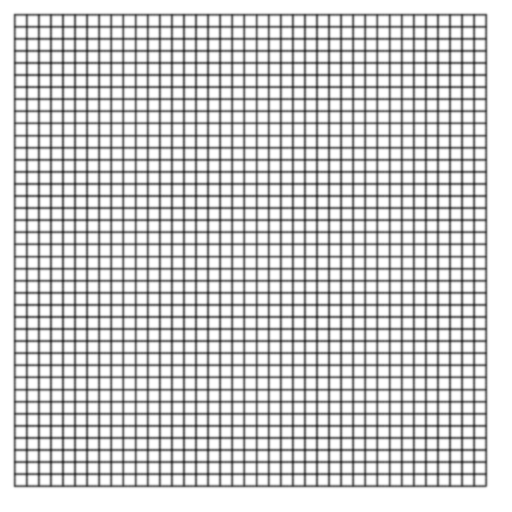}
\includegraphics[width=0.24\textwidth]{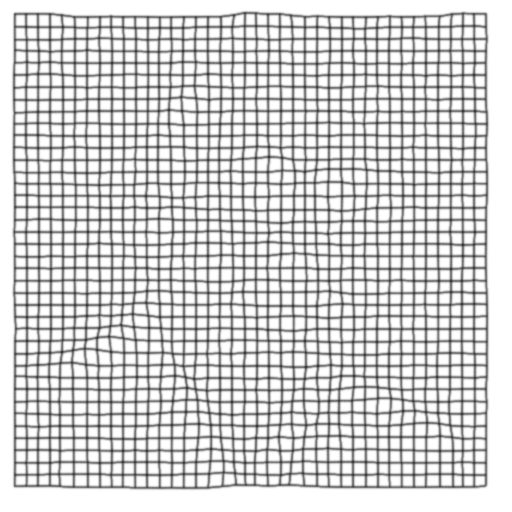}
\includegraphics[width=0.24\textwidth]{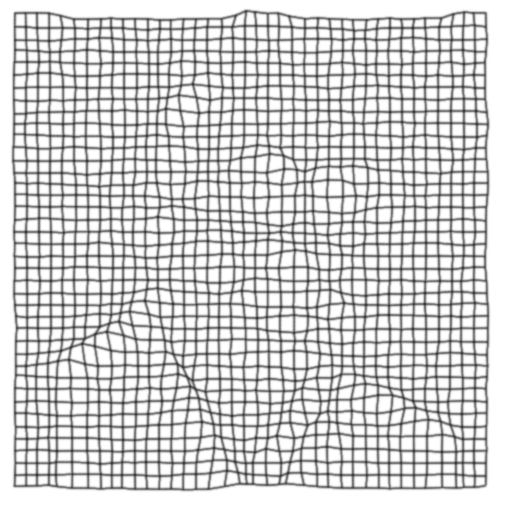}
\includegraphics[width=0.24\textwidth]{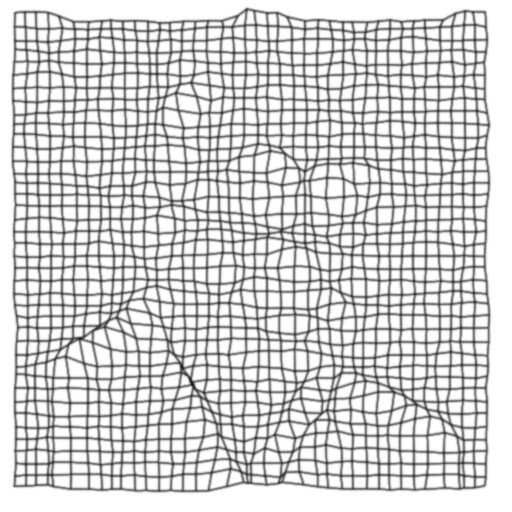}
\caption{{\em Approximating Zel'dovich}: a discretised displacement field based on an image of Y. Zel'dovich plus gaussian noise, multiplying the displacement with increasing factors from left to right, starting at zero.}\label{fig:face}
\end{figure}

A displacement field can be regarded as coordinate transformation from some (virtual) coordinate system with a constant charge per coordinate volume to a (physical) coordinate system with a perturbed density (of charges, mass, etc.). When the perturbations are small, and the unperturbed quantity is denoted by an overbar, the coordinate transformation to obtain the displacement field that is associated with some density field $\rho(\vec x) = \bar \rho (1+ D(\vec x))$ can be defined by,
\begin{align}
\bar \rho (1+D(\vec x)) d^{N-1}x =&  \bar \rho  d^{N-1} x' ,\\
(1+D(\vec x)) =&  \left| \frac{\partial x^{i}}{\partial x^{j'}} \right| ,
\end{align}
such that an observer at fixed $\vec x'$ is co-moving with the charges $\rho$. 
For small $D(x^k)$ and in absence of vorticity, the coordinate transformation is solved by 
\begin{align}
x^{i} =& x^{i'} -  \frac{\vec \nabla' }{{\nabla'}^2} D(\vec x') \nonumber\\
 =& x^{i'} -  \frac{\vec \nabla }{\nabla^2} D(\vec x) + \mathcal{O}(D^2),
\end{align}
where ${\nabla'}^2 \equiv \sum_{i'} \partial^2_{i'}$ and 
\begin{align}
\frac{1}{\nabla^2}D(\vec x)=\int \frac{d^{N-1}k\,d^{N-1}\tilde x}{\left(2\pi\right)^{N-1}} \,\, k^{-2}  D\left(\vec {\tilde x}\right)\,\, e^{i\vec k\cdot\left(\vec x - \vec{ \tilde x  }\right)} 
\end{align}
This relation is independent of the theory that gives rise to the density field\footnote{A similar conclusion is drawn in Ref.~\cite{Joyce:2004qz}, based on the continuity equation such that velocities {\em are} in agreement with a physical theory.}, as exemplified in Fig.~\ref{fig:face}. However, since we set vorticity to zero, the Poisson equation that determines the Newtonian potential is equal to the equation that gives the deformation tensor $\partial x^i/\partial x^{j'}$. One can identify the scalar $D(\vec x)$ with the Newtonian potential, and obtain the Zel'dovich Approximation~\cite{Novikov:2010ta, Zeldovich:1969sb}.

In the $x'$ coordinate system, the equal-weight particles are not displaced, such that this is a Lagrangian coordinate system\footnote{In cosmology, the labels $\vec x'$ are often referred to as the initial positions of particles at the past singularity. This is not a consequence of Lagrangian perturbation theory, but this is a consequence of the fact that in cosmology usually only growing modes are considered, such that $D(x) \rightarrow 0$ at past infinity. Fundamentally, it is not necessary for $\vec x'$ and $\vec x$ to coincide at any time, especially since in the context here there is no reference to time at all.}: the coordinates are at rest with the charges, which do not necessarily have to be masses.

If the density field has a continuous time dependence, $\rho(\vec x) \rightarrow \rho(\tau, \vec x)$, the velocities $\vec v^{(c)}$ of the virtual equal-weight particles are simply given by the time derivative of the field,
\begin{align}
\vec v^{(c)}(\tau, \vec y) =  -  \frac{\vec \nabla }{\nabla^2}\partial_t \rho(\tau, \vec x).\label{eq:pseudoVelField}
\end{align}

When one is dealing with a curved manifold, care needs to be taken with the meaning of the displacement and density fields. For the displacements to correspond to the coordinate positions of particles on an arbitrary manifold, the density field used for the coordinate transformation must be the `bare' density of particles in coordinate space, not taking into account the curvature of space. If one were to take the proper density perturbations as the generator for the coordinate transformation, one would obtain the displacement of particles in a (purely spatially) transformed coordinate gauge, in which the trace of the perturbations of the spatial metric vanishes.

Instead of employing the purely spatial density--displacement duality above, one may perform a similar coordinate transformation using the proper energy density on the spacetime,
\begin{align}
\rho(x)\sqrt{-g}d^Nx = \bar\rho(1+D(x))\sqrt{-g}d^Nx = \bar\rho\sqrt{-g'}d^Nx',
\end{align}
in which case the $x'$ coordinate system acquires the meaning of a synchronous comoving gauge, associated with the species of $\rho$, as explained in~\cite{Rampf:2013ewa,Rigopoulos:2013nda, Rigopoulos:2014rqa}. Such a transformation makes explicit where General Relativity comes into play compared to Newtonian dynamics, but is not useful for the purpose of this work.

When a problem is discretised {  (see Appendix~\ref{sec:discrete})}, the Eulerian description amounts to following a density field on a regular grid in $\vec x$ at positions $\vec x_{ijk}$, while the Lagrangian description amounts to following a displacement field for points on a regular grid in $\vec x'$, which translates to a curved mesh in $\vec x$-space. Here `regular' does not necessarily mean Cartesian, because for example glass initial conditions~\cite{White:1994bn} or alternatives~\cite{Hansen:2006hj} take an irregular partitioning of $\vec x'$-space.

In general, at any time $dM(\tau, \vec x) = \bar\rho d^3x' = \bar\rho \left| \frac{\partial x^{i'}}{\partial x^{j}} \right| d^3x$. This expression continues to hold in situations where $\vec x'(\vec x)$ is not single valued and $\delta(x) \geq \mathcal{O}(1)$, {\em i.e.} the phase-space sheet has folded~\cite{Abel:2011ui}. It must be noted that its evaluation becomes nontrivial then.

\section{Linear displacement and velocity fields in metric theories of Gravity\label{sec:mainIC}} 
\subsection{Geometry}
The next section applies to any metric theory of gravity, with a covariant derivative defined by,
\begin{align}
	\nabla_\mu v^\nu =& \partial_\mu v^\nu + \Gamma^{\nu}_{\mu\lambda}v^\lambda,\\
	\Gamma^{\nu}_{\mu\lambda} =& \tfrac{1}{2} g^{\nu\alpha}\left(\partial_{\mu}g_{\lambda\alpha} + \partial_{\lambda} g_{\alpha\mu} - \partial_{\lambda} g_{\mu\alpha} \right),
\end{align}
with the Christoffel connection $\Gamma^{\nu}_{\mu\lambda}$. We do not refer to the equations that source the metric with a matter configuration. That is, we define all modifications of gravity as those that change the Einstein equations and / or those that can be written as additional matter content inside the energy-momentum tensor.

\subsection{Hydrodynamics}
At the linear level and hence at the Cauchy surface of N-body simulations, the cosmic density fields under consideration can be described as fluids.
Let us hence first layout the definitions for relativistic hydrodynamics. All contents of this section can be traced back to Refs.~\cite{1959flme.book.....L, Kodama:1985bj, Ma:1995ey, Sawicki:2012re}.

This paper focusses on dynamics entirely attributable to scalar quantities. All is perturbed about a background Friedmann-Lema\^itre-Tolman-Bondi solution, denoted by overbars, such that background quantities have only time dependence while perturbed quantities (indicated by $\Delta$) have time and space dependence. Hereafter we drop the time and space dependence in most functions. 
For a fluid with 
four-velocity $U^{\mu} = dx^\mu / \sqrt{-ds^2} $, $U^\mu U_\mu = -1$, 
define the transverse projector\footnote{The symbol $\perp$ can be pronounced `perp', for `perpendicular'.},
\begin{align}
\perp_{\mu\nu}=g_{\mu\nu}+U_{\mu}U_{\nu},
\end{align}
which projects into the plane orthogonal to $U^{\mu}$, the spatial slices for an observer comoving with the fluid. The unperturbed $\bar U^i=0$, and $\partial_i \delta U^i \equiv \theta$.

The energy momentum tensor $T_{\mu\nu}$ then carries the following information,
\begin{itemize}
\item energy density $\rho = \bar\rho + \delta \rho  = \bar\rho(1+\Delta_\rho) \equiv U^{\mu}U^{\nu}T_{\mu\nu}$, 
\item pressure $P = \bar P +\delta P= \bar P(1+\Delta_P)\equiv \tfrac{1}{3}\perp^{\mu\nu}T_{\mu\nu}$, 
\item energy flow (or heat transfer) $q^{\mu} \equiv \perp^{\mu\nu}U^{\lambda}T_{\nu\lambda}$,
\item anisotropic shear perturbation $\Sigma^{\mu\nu}$, 
\end{itemize}
and is decomposed as
\begin{align}
	T^{\mu}_{{\phantom{\mu}}\nu} =&  \rho U^{\mu} U_{\nu} + P \perp^{\mu}_{{\phantom{\mu}}\nu} + U^{\mu} q_{\nu} + U_{\nu} q^{\mu}+ \Sigma^{\mu}_{{\phantom{\mu}}\nu}.
\end{align}
Owing to the scalar nature of the system one can further simplify with
\begin{align}
q^\mu=&-a(\tau)\left(\rho + P\right)\perp^{\mu\nu}\frac{\nabla_{\nu}}{\nabla^2}q,\\
 \Sigma^{\mu\nu}=& -\tfrac{3}{2}a(\tau)^2\left(\rho + P\right)\left(\perp_{\mu\lambda}\perp_{\nu\alpha}-\tfrac{1}{3}\perp^{\mu\nu}\perp^{\lambda\alpha}\right) \frac{\nabla_{\lambda}\nabla_{\alpha}}{\nabla^2}\sigma, 
\end{align}
where we chose pre-factors for convenience, such that $\sigma$ corresponds to the definition in Ma \& Bertschinger~\cite{Ma:1995ey} when $q=0$.

In an arbitrary gauge, the scalar part of the metric can be written as~\cite{Bardeen:1980kt},
\begin{align}
\frac{ds^2}{a(\tau)^2}=-(1+2A) d\tau^2 - 2 B_i d\tau\,dx^i + \left[(1+2H_L)\eta_{ij} + 2h^T_{ij}  \right]dx^idx^j ,
\end{align}
where $\eta_{\mu\nu}$ is the Minkowsky metric, and
\begin{align}
B_i =& \int \frac{d^3k}{\left(2\pi\right)^3}\frac{k_i}{k}B_{{\vec k}} e^{i\vec k \cdot \vec x},\\
h^T_{ij} =& \left[\frac{\partial_i\partial_j}{\nabla^2} - \tfrac{1}{3}\eta_{ij}\right]H_T,
\end{align}
where all scalar potentials $A$, $B$, $H_L$ and $H_T$ are small and spacetime dependent\footnote{Compare with the convention in \cite{Ma:1995ey}, in the longitudinal gauge we have ($A=\psi,H_L=-\phi$) and in the synchronous gauge ($6H_L=h,H_T=6\eta+h$).}.

Any number density in the frame of an observer comoving with the fluid at velocity $U^{\mu}$ is a scalar $n=\bar n + \delta n = \bar n(1 + \Delta_n)$, and the number transport is $n^{\mu} = - n U^{\mu}$. If the number is conserved, we have at the linear level and independent of whether the fluid's energy is conserved,
\begin{align}\hspace{2cm}
	\nabla_\mu\, n^{\mu} =& \, 0, \hspace{2cm} \mbox{[number conservation]}\\
	 \bar n \propto & a^{-3}, \\ 
	\dot \Delta_n =& -(\theta + 3 \dot H_L),\label{eq:ParticleConservation} \hspace{1.63cm} \mbox{[linear order]}
\end{align}
where an overdot denotes a derivative with respect to conformal time $\tau$. The number can be associated with microscopic particles, but just as well with macroscopic simulation vertices (also often referred to as `particles'). 

The energy conservation equation $\nabla_\mu T^{\mu\nu} = 0$ corresponds to the continuity and Euler equations, at linear level in Fourier space\footnote{See Ref.~\cite{Hu:2004xd} for the special case of the perfect fluid, {\em i.e.} $q=\sigma=0$.}
\begin{align}
	\dot{\bar\rho} =& - 3\mathcal{H}\left(\bar\rho + \bar P\right),\\
\dot\Delta_\rho =&  \left(1 + w\right)\left( q - \theta - 3\dot  H_L\right) + 3\mathcal{H}\left(w - c_s^2\right)\Delta_\rho,\label{eq:energyConservation}\\
\dot \theta + k \dot B - \dot q=&  k^2 A  +k^2 \Delta_P   - \mathcal{H} (1-3w) (\theta + kB - q)-\frac{\dot w }{w+1}(\theta + kB - q) - k^2\sigma,\label{eq:velocityEquation}
\end{align}
where $\mathcal{H}\equiv \dot a(\tau)/a(\tau)$ and $w(\tau)\equiv\frac{\bar P}{\bar \rho}$. 
Note that at the linear level, the continuity equation is only sensitive to the trace of the spatial perturbations of the metric, $3H_L$. The scalar potentials $A$ and $B$ only affect the Euler equation for the velocities, while the transverse traceless potential $H_T$ does not enter the energy conservation equations at the linear level.

\subsection{Constant or variable mass per vertex during the fluid phase}
The phase-space of a fluid is discretised for the purpose of a numerical simulation, upon the start of which a particle picture may be followed, which goes beyond the fluid description when perturbations become nonlinear and trajectories start crossing. This means that a mesh is laid out in real space with a velocity associated to each vertex of the mesh, the `particles'. {  For a summary of the existing methods for discretisation, see Appendix~\ref{sec:discrete}.}

Although this it is not a fundamental condition, all numerical cosmological simulations known to the authors are based on a fixed number of vertices in phase-space (particles) which is preserved throughout the simulation. This is not to be confused with the Poisson solver, which during the simulation at any time can choose an adaptive mesh to approximate the gravitational potential(s) at the desired resolution, provided the positions of the fixed number of particles.

If the vertices of the fluid are comoving with the velocity field of the fluid, then their proper number density follows Eq.~\eqref{eq:ParticleConservation}.
Comparing Eq.~\eqref{eq:ParticleConservation} to Eq.~\eqref{eq:energyConservation}, it is clear that the special case of a fluid with constant equation of state $\dot w=0$,  constant sound speed $w=c_s^2$ which for this case is equal to $c_s^2=\delta P /\delta\rho$, and zero energy flow $q=0$, can be modelled discretely by only accounting for the positions and velocities of vertices, associating an energy density to each vertex with, 
\begin{align}
\Delta_\rho = (1+w) \Delta_n. \hspace{2cm} \mbox{[$\dot w = q = 0$, $w=c_s^2$, $\nabla_{\mu} T^{\mu\nu}=0$ ]}
\end{align}
Note that the fluid need {\em not} be perfect, as $\sigma$ need not be zero for this condition, which is why the statement above may apply to other fluids than dust. {  The quantity $\Delta_\rho$ is to be evaluated on the discrete vertices on which initial conditions are generated.}

If any of the three conditions is broken during the time span of the simulation, one needs to model the dynamical energy density of the vertices. In other words, the `particles' do not have a `constant mass', and energy density and pressure at each vertex are given by equating $\rho(\tau, \vec x) = m(\tau, \vec x) n(\tau, \vec x)$ and $p(\tau, \vec x) = n(\tau, \vec x)T(\tau, \vec x) $, where $m$ can be thought of as mass and $T$ as temperature, although strictly speaking they can have other meanings,
\begin{align}
	m =& \frac{\bar \rho}{\bar n}\left( 1 + \Delta_\rho - \Delta_n \right), \\
	T =& \frac{\bar P}{\bar n}\left( 1 + \Delta_P - \Delta_n \right) = \frac{\bar \rho}{\bar n}\left(w + \frac{\delta P}{\delta\rho}\Delta_\rho - w\Delta_n\right).
\end{align}
{  It should be clear now that $m$ and $T$ are evaluated at the discrete set of vertices (`particles').}
One is free to set for example $\frac{\bar \rho}{\bar n} = 1$ at some given time. Moreover, depending on the extent to which a gauge is fixed, one may have further freedom to set $\Delta_m = 0$ at a convenient time of choice, which amounts to different choices of cutting the density field into particles. 

\subsection{Bare densities and velocities\label{sec:bare}}
A simulation acts in a coordinate space. The bare density of vertices is at any time step given by the number of vertices inside a given volume in coordinate space~\cite{Adamek:2013wja}. This is related to the proper density by,
\begin{align}
	n^{\bare} d^3x = n \sqrt{g^{(3)}} d^3x,
\end{align}
where $g^{(3)} = \det g_{ij} = a(\tau)^6(1+6H_L) + \mathcal{O}(\epsilon^2)$, the determinant of the spatial part of the metric, the three-metric, such that\footnote{As noted in~\cite{Adamek:2013wja}, for simulations in the longitudinal gauge, $n^{\bare}_{\rm (long)}$ corresponds to $n_{\rm (comov)}\equiv n_{\rm (comov)}^\bare / \sqrt{g_{\rm (comov)}^{(3)}} $ of the comoving gauge, but this coincidence does not occur for other gauges in which the transverse traceless perturbation of the metric does not vanish.} 
\begin{align}
	\Delta_n = \Delta_n^{\bare}  - 3H_L.\label{eq:dnbare}
\end{align}
Thus, the displacement fields are expressed in terms of coordinates (and not proper distances), such that they must be generated in coordinate space based on the bare densities. Density fields, on the other hand, continue to express proper densities and need no notion of bare density.

A displacement field based on number density $n$, must be consistent with the theory at any time step. This implies that the motion along a displacement field at different time steps must reproduce the correct velocities. Indeed, we find,
\begin{align}
	\theta^{\bare} = -\dot \Delta_n^{\bare} = -\dot\Delta_n - 3\dot H_L = \theta, \label{eq:baredot}
\end{align}
as per Eq.~\eqref{eq:pseudoVelField}~and~Eq.~\eqref{eq:ParticleConservation}.

Note that the definitions of mass and temperature are unaffected by the notion of bare density,
\begin{align}
	m =& \frac{\bar \rho}{\bar n}\left( 1 + \Delta_\rho - \Delta_n \right) =  \frac{\bar \rho}{\bar n}\left( 1 + \Delta_\rho^{\bare} - \Delta_n^{\bare} \right), \\
	T =& \frac{\bar P}{\bar n}\left( 1 + \Delta_P - \Delta_n \right) =  \frac{\bar P}{\bar n}\left( 1 + \Delta_P^{\bare} - \Delta_{n}^{\bare} \right) .
\end{align}

In summary, particle positions in a {\em relativistic} simulation are obtained from applying the density--distance duality to bare number densities, while particles masses and internal energies are set through the equations above. {  Again, these quantities are to be evaluated on the discrete positions of the simulation particles.}

\subsection{Adiabatic and isocurvature initial conditions for multiple fluids and fields}

Now that we have shown in brief how to relate a solution to the stress-energy tensor to a displacement field for a quantity described in a Lagrangian picture, in arbitrary coordinates, we can address the problem of discretising Lagrangian quantities and Eulerian quantities simultaneously, even though the perturbations are strictly at different coordinates. The coordinates $\{t, \vec x\}$ are the coordinates of the simulation. An Eulerian quantity's phase space is sampled and simulated on a regular grid in $\vec x$-space, {\em i.e.} on $\vec x_{ijk}$, with indices $\{i,j,k\}$ labeling discrete positions in dimensions $\{x^1, x^2, x^3\}$ respectively. Tracers of the phase space of a Lagrangian quantity are on a regular grid in that quantity's homogeneous-coordinate-density system (``equal-charge tracers''), $\vec y_{ijk}$, with indices $\{i,j,k\}$ labeling discrete positions in dimensions $\{y^1, y^2, y^3\}$. The coordinates are related by $\vec y = \vec x +  \vec \coordDisp$, such that a regular grid in one space becomes a curved mesh in the other; the grid points in both spaces do not coincide.

How does one generate realisations of initial conditions in real space, when multiple grids are present, as many as there are Lagrangian quantities plus one for all Eulerian quantities? The short answer is: { the same random numbers can be used on all grids}, as long as displacements are sufficiently small. In a more General Relativistic language: as seeds of linear perturbations, { the random numbers are gauge independent { ($\vec \xi$ in Appendix~\ref{sec:discrete})}}. This follows straightforwardly from the argument in section~\ref{sec:anyPlainDisp}, which can be summarized by $\vec x_{ijk} = \vec y_{ijk} +  \vec \coordDisp(\vec y_{ijk}) =  \vec y_{ijk} +  \vec \coordDisp(\vec x_{ijk}) + \mathcal{O}(\epsilon^2) $, where $\epsilon$ denotes all quantities that are assumed small, being displacements, potentials, densities and so on.

By virtue of the closed set of linear differential equations that describes the system in the linear regime,
\begin{align}
{\mathbf M}\left(\tau, \vec k\right) {\mathbf f}_{\vec k}(\tau) = 0,\label{eq:genericODE}
\end{align}
where $\mathbf{M}\left(\tau, \vec k\right)$ denotes a matrix with differential operators in $t$ and nonlinear functions in $\vec k$, and ${\mathbf f}_{\vec k}(\tau)$ denotes the set of dynamic degrees of freedom, all random input in a realisation of initial conditions is encoded in the initial conditions of the differential equations. That is, the differential equations are deterministic, and the solution changes linearly with the initial conditions,
\begin{align}
{\mathbf f}_{\vec k}(\tau) = \left(\begin{array}{c} D_1 \left(\tau, \vec k\right) f^{\rm ini}_{1,\vec k} \\ \ldots \\ \ldots \\ D_m \left(\tau, \vec k\right) f^{\rm ini}_{m,\vec k} \end{array} \right),
\end{align}
where ${\mathbf D}\left(\tau, \vec k\right)$ solves the system of equations~\eqref{eq:genericODE}.
Generating one realisation at any time in cosmic history, amounts to choosing random numbers ${\mathbf f}^{\rm ini}_{\vec k}$ that in the real universe actually are drawn at the hot big bang, for example by the inflaton as quantum correlations decohere and become classical. { The quantity ${\mathbf f}^{\rm ini}_{\vec k}$ can be identified with $\vec \xi$ in Appendix~\ref{sec:discrete}.}

A realisation is entirely described by its Fourier transform, regardless of whether the random numbers are drawn in Fourier space or in real space (as in \cite{Bertschinger:2001ng}{ , see Appendix~\ref{sec:discrete}}), and regardless of whether the distribution is gaussian or not. All quantities in the system are related linearly by functions of wavenumber $\vec k$, because we consider the system at the linear regime. In practice, the Fourier transform of the perturbations in each quantity are a time and space dependent amplitude, multiplied by a normal Gaussian random number (with or without correlations, depending on the type of initial conditions).

In summary, in the case of adiabatic initial conditions, all quantities in the vector ${\mathbf f}_{\vec k}(\tau)$ share the same random seed, 
\begin{align}
{\mathbf f}_{\vec k}(\tau) = \left(\begin{array}{c} D_1 \left(\tau, \vec k\right) \\ \ldots \\ D_m \left(\tau, \vec k\right)\end{array} \right) f^{\rm ini}_{\vec k} , \hspace{2cm} \mbox{[adiabatic]}
\end{align}
regardless of whether these quantities describe a lagrangian displacement field in a mesh which appears curved in Eulerian coordinates, or whether these quantities are on a regular Eulerian grid. Already in a $\Lambda$CDM universe, baryons and Cold Dark Matter have slightly different transfer functions, which should be taken into account when generating their initial conditions based on the same random seed, as first applied in~\cite{Bird:2010mp}. Owing to the linearity of the system, any type of isocurvature perturbations can then be expressed as,
\begin{align}
{\mathbf f}_{\vec k}(\tau) = \left(\begin{array}{c} D_1 \left(\tau, \vec k\right) \\  \ldots \\ D_m \left(\tau, \vec k\right)\end{array} \right) f^{\rm(1), ini}_{\vec k} + \left(\begin{array}{c} D_1 \left(\tau, \vec k\right) \\ \ldots  \\ D_m \left(\tau, \vec k\right)\end{array} \right) f^{(2),\rm ini}_{\vec k} , \hspace{2cm} \mbox{[adiabatic + isocurvature]}
\end{align}
and so on.
These statements hold for any type of initial seed, whether it is gaussian or not. Notably, in~\cite{Burrage:2015lla} it is pointed out that exactly a scalar modification of gravity may acquire its own spectrum of perturbations, giving rise to isocurvature perturbations.

\section{Non-relativistic initial conditions\label{sec:nrlim}}
\subsection{Non-relativistic limit}
The non-relativistic limit for perturbations in an expanding universe is obtained when velocities are small compared to the speed of light. Furthermore, the Newtonian limit is obtained when pressure is small, in which case the linearised perturbations of the fluid are described by~\cite{1980lssu.book.....P},
\begin{align}
	\dot \Delta_{\rho} =& - \theta + q, \hspace{2cm} \mbox{[Newtonian, $\sigma=0$, $w=0$, $c_s^2\ll 1$]}\\
	\dot \theta - \dot q=&  k^2 \phi  + c_s^2 k^2 \Delta_\rho - \mathcal{H} (\theta - q),
\end{align}
where we continue employing the usual co-expanding\footnote{Normally called comoving coordinates, we exceptionally say co-expanding in order to avoid confusion with a coordinate gauge for relativistic perturbations, such as the comoving gauge.} coordinate system with conformal time. However, these equations neglect the effect of pressure on the energy density present in relativity (associated to the work needed to compress a fluid with pressure). The equations can be modified to obtain the non-relativistic limit that has the correct continuation to the General Relativistic equations~\cite{Lima:1996at}, which in practice amounts to removing all references to geometry and installing a single gravitational potential,
\begin{align}
	\dot \Delta_{\rho} =& (1+w) (q - \theta), \hspace{2cm} \mbox{[Non-relativistic, $\sigma=0$, $w=c_s^2\neq0$, $\dot w = 0$]}\label{eq:nreom}\\
	\dot \theta - \dot q=&  k^2 \phi  +\frac{c_s^2 }{ (w+1)}k^2 \Delta_\rho - \mathcal{H}  (\theta  - q). \label{eq:nreomtheta}
\end{align}
The Newtonian limit is necessary for setting initial conditions for Newtonian simulations. Power spectra for the distribution of densities, velocities and potentials are however readily obtainable from solvers of relativistic Boltzmann equations, such as {\sc Class}~\cite{Blas:2011rf} and {\sc Camb}~\cite{Lewis:1999bs}.

One could use the gauge freedom to fix the gauge such that Eqs.~(\ref{eq:nreom}, \ref{eq:nreomtheta}) hold at all scales~\cite{Flender:2012nq, Rampf:2013dxa}. We briefly elaborate on that in Section~\ref{sec:gauges}.  In this subsection we focus however on the weak field limit, in which initial conditions for Newtonian N-body simulations are inevitably set. An approximate translation from Newtonian large-scale results to relativistic interpretations can be found in ~\cite{Green:2011wc}.

The proof that Newtonian gravity is the weak field limit of General Relativity is textbook material, where it is usually shown that the equations of motion in the longitudinal gauge ($H_T=B=0$) reproduce the non-relativistic equations of motion~\eqref{eq:nreom}, although the same can be proven in the comoving gauges ($\theta=B$)~\cite{1980lssu.book.....P}. The density perturbations of both these gauges, computed in terms of quantities of an arbitrary gauge, are~\cite{Bardeen:1980kt},
\begin{align}
\Delta_\rho^{\comoving} =& \Delta_\rho + 3(1+w)\frac{\mathcal{H}}{k}\left(\theta - B\right),\\
\Delta_\rho^{\longitudinal} =& \Delta_\rho + 3(1+w)\frac{\mathcal{H}}{k}\left(\frac{\dot H_T}{k} - B\right).
\end{align}
The pre-factor $\mathcal{H}/k$ is by definition a measure of the smallness of velocities in the problem\footnote{In the Newtonian picture, cosmic expansion is a recession velocity $v_{\rm recession} = \mathcal{H}\, d$, for arbitrary distance $d$. Identifying $k$ with a wavelength $\lambda = 2\pi/k$, we have $v_{\rm recession}=1$ for $k=\mathcal{H} / (2\pi)$, hence for simulations that include scales close to the Hubble radius. Thus, the Newtonian limit is valid for simulations of boxes smaller than $2\pi/(3(1+w)\mathcal{H})$~\cite{Rigopoulos:2013nda}.}, such that $\mathcal{H}/k \rightarrow 0$ gives the lowest order terms in the Newtonian limit. In other words, the density perturbation computed in {\em any} gauge reduces to the same density perturbation in the non-relativistic limit, and is a solution to the set Eqs.~(\ref{eq:nreom}, \ref{eq:nreomtheta}),
\begin{align}
	\Delta_\rho = \Delta_\rho^{\comoving} + \mathcal{O}(\tfrac{v}{c}) = \Delta_\rho^{\longitudinal} + \mathcal{O}(\tfrac{v}{c}).
\end{align}
This means that for the non-relativistic limit, one can use the density perturbations computed in {\em any} gauge (also gauges in which $\theta=0$), and set the velocities by,
\begin{align}
\theta^{\newt} =& \frac{-\dot \Delta_\rho}{1+w}. \hspace{2cm} \left [\frac{3(1+w)\mathcal{H}}{k} \ll 1 \right]
\end{align}
More generally, this equation holds under any extension of general relativity (modified theory of gravity), provided that the fluid in question satisfies $\dot \Delta_{\rho} = -(1+w) \theta$ with $\dot w=0$.

To summarise: 
\begin{itemize}
\item for a newtonian simulation, positions are obtained from the density--displacement duality applied to $\Delta_\rho$, and velocities from $\dot\Delta_\rho$,
\item while for a relativistic simulation,  the density--displacement duality is applied to bare quantities, $\Delta_n^{\bare}$ and $\dot\Delta_n^{\bare}=-\theta$. 
\end{itemize}

\subsection{Gauges and the Newtonian limit\label{sec:gauges}}
From comparing Eqs.~(\ref{eq:energyConservation},~\ref{eq:velocityEquation}) to Eqs.~(\ref{eq:nreom}, \ref{eq:nreomtheta}), it is obvious that a gauge in which $H_L=0$ reproduces the same linear equations in General Relativity as in Newtonian gravity.~\footnote{During the final stages of preparation of this manuscript, the same was pointed out in Ref.~\cite{Fidler:2015npa}, where it is shown that for Cold Dark Matter, the gravitational potential follows the Newtonian Poisson equation, such that the {\em linear} dynamics look completely Newtonian at relativistic scales.} It is tempting to believe that Newtonian simulations hence can be used on relativistic distances $d$ such that $d\,\mathcal{H} = \mathcal{O}(1)$. One strong hint that this is the case, comes from the fact that linear perturbation theory agrees well with observations of large scale structure in the universe, suggesting that small-scale nonlinearities do not affect the large scale dynamics. The growth of structure is hierarchical: small scales enter the nonlinear regime first, while large scales continue to evolve linearly. Moreover, all modes follow a sequence of (1) relativistic dynamics (super Hubble), (2) newton linear dynamics, (3) nonlinear dynamics~\cite{Rigopoulos:2013nda}.  While this suggests that long wavelength modes in Newtonian N-body simulations are solved for properly, this does not at all mean that the full nonlinear dynamics of the short wavelength modes are the same for relativistic and Newtonian systems. What is needed is a proof that the fully nonlinear equations of Newtonian gravity and General Relativity agree, even when taking into account all scales (since obviously, when any scale goes nonlinear, formally the entire system is nonlinear). It is often claimed that even the nonlinear density solutions only source the relativistic potentials at the linear level, but as pointed out in~\cite{Green:2011wc}, it is inconsistent to linearise the left-hand side (the gravity part) of the Einstein equations without linearising the right-hand side (the matter part). When the matter part contains nonlinear contributions, the average of the matter density will no longer agree with the definition of the background matter density, around which perturbation theory was setup. Properly taking this mismatch into account, amounts to the Buchert formalism (see Ref.~\cite{Buchert:2011sx} for a review), which schematically can be summarised as,
\begin{align}
	\bar G_{\mu\nu} ( \bar g_{\mu\nu}(\tau) ) =& 	\bar T_{\mu\nu} ( \bar \rho(\tau), \bar P(\tau),  \ldots ), \\
	\delta G_{\mu\nu} ( \delta g_{\mu\nu}(\tau, \vec x) ) =& 	\delta T_{\mu\nu} ( \delta \rho(\tau, \vec x), \delta P(\tau, \vec x),  \ldots ),& \mbox{[ordinary perturbation theory]}\\
	\left<G_{\mu\nu} ( g^{\rm fully\,nonlinear}_{\mu\nu}(\tau, \vec x) )  \right> \neq& \bar T_{\mu\nu} ( \bar \rho(\tau), \bar P(\tau),  \ldots ) , &\mbox{[averaging vs. background]}
\end{align}
where both the left and right-hand side of the last line are pure functions of $\tau$, but different pure functions of $\tau$.

In General Relativity, intuition tells that wavelengths larger than the scales considered, can be included as a locally constant spatial curvature term (the universe locally is open or closed). However, this only addresses the density contributions, while the Einstein equations contain higher order gradients in the potential as well. Under general modifications of gravity, Birkhoff's theorem no longer holds, and the above intuition is flawed.

In summary, it is not obvious that two perturbation theories that are identical in their linear limits, are identical in their full nonlinear regimes.

 In~\cite{Adamek:2015gsa} is was found numerically that the mismatch between average and background quantities may be tiny, but a rigorous mathematical proof of the agreement between fully nonlinear Newtonian gravity and General Relativity in cosmic structure formation does currently not exist.

\section{Scale dependent growth in the quasi-static approximation\label{sec:qsa}}
\setlength{\unitlength}{0.39375\textwidth}
\begin{figure}
\begin{center}
\includegraphics[width=0.7\textwidth]{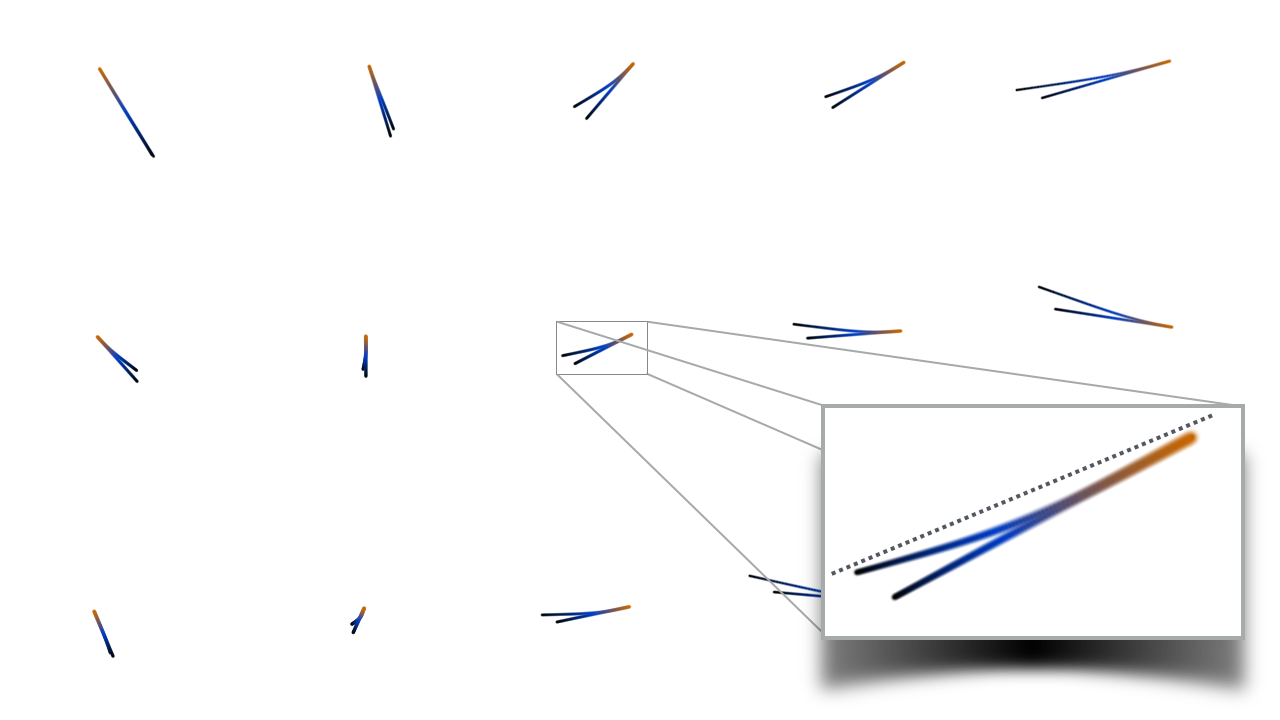}
\includegraphics[height=0.39375\textwidth]{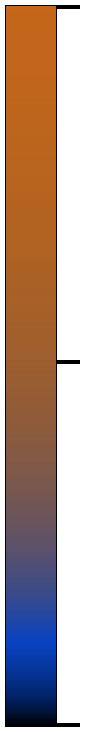}
\begin{picture}(0.2,1)
\put(0, 0) { $z = 10$ }
\put(0, 0.49) { $z = 55$ }
\put(0, 0.98) { $z = 100$ }
\end{picture}
\end{center}
\caption{A 3x5 of patch of a $128^2$ slice of a realization $128^3$ vertices (particles) of a Dark Matter density field at a comoving size of 200 Mpc (1.56 Mpc inter-vertex distance), at newtonian displacements varying from redshift $z=100$ down to $z=10$, in a cosmology endowed with designer $f(R)$ gravity in $w$CDM background ($w=-0.95$) with $B_0=0.9$, compared to a plain $\Lambda$CDM cosmology. The $f(R)$ trajectories can be distinguished from the $\Lambda$CDM counter parts, by their curved shape, as emphasised in the inset. The straight dashed line connecting start and end points serves as a guide to the eye, to recognise the curved shape of the trajectory. This displacement field is linear, yet the particles do not follow straight lines. }\label{fig:curves}
\end{figure}
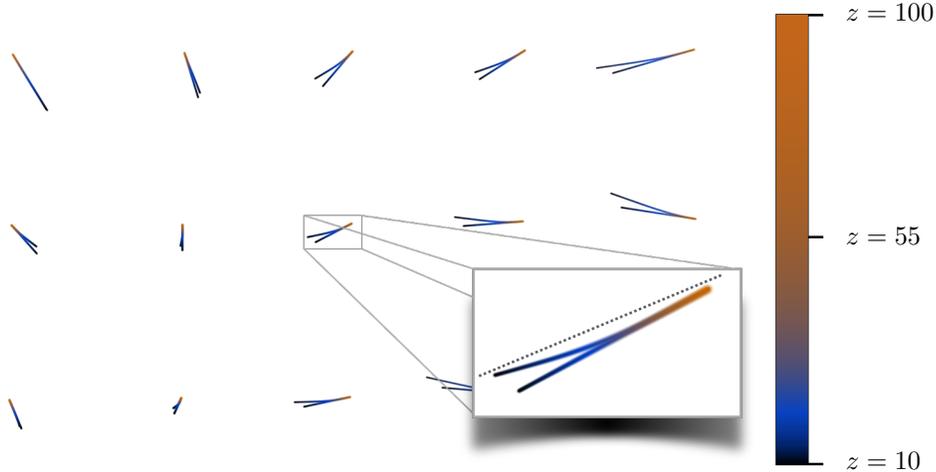

In this section, we show that for Dark Energy and Modified Gravity models (in short DE/MG models) the velocities of Dark Matter particles are scale dependent in linear perturbation theory, already in the quasi-static approximation. Scale dependence of the density transfer function of Dark Matter in Fourier space generally translates into a varying direction in real space. That is, the dust particles do not move on straight lines. Here we choose to apply the quasi-static approximation in order to make an intuitive transition from General Relativity to the more general Effective Field Theory of Modified Gravity. In the next section we abandon the quasi-static approximation\footnote{In our numerical calculation, we do not assume the quasi-static approximation, and evolve the full dynamical equation of motion of the extra scalar field.}.

We fix the metric to the longitudinal gauge, $B=H_T = 0$, $A=\Psi$ and $H_L=-\Phi$,
\begin{align}
  ds^2=a^2(\tau)\left[-(1+2\Psi)d\tau^2+(1-2\Phi)d\vec x^2\right]\;.
\end{align}
Via the quasi-static approximation, we can parametrise the Poisson and anisotropic stress equations into the following generic form \cite{Bean:2010zq}:
\begin{align}
  -\nabla^2_x\Phi(\tau,\vec x) &= 4\pi G a^2Q(\tau,\vec x)\bar\rho_i\Delta_i(\tau,\vec x)\;,\label{modfdPoisson}\\
  -\nabla^2_x(\Psi-R\Phi)&=12\pi Ga^2\bar\rho_i(1+w_i)\sigma_i Q(\tau,\vec x)\label{gamma}\;,
\end{align}
where $Q(\tau,\vec x)$ and $R(\tau,\vec x)$ describe the variation of Newton's constant in the environment and the anisotropic stress tensor induced by the modification of gravity, respectively. The quasi-static approximation was proposed via ($\mu,\gamma$) functions in \cite{Bertschinger:2008zb}, but mixing the gravitational shear and the anisotropic stress from relativistic species\footnote{In the only CDM+modified gravity case, the functions $(\mu,\gamma)$ are related to $(Q,R)$ by: $Q=\mu\gamma\;,\;\; R=\gamma^{-1}\;.$}. Several phenomenological parameterisations of these functions exist in the literature. We refer to the Refs. \cite{Brax:2012gr,Silvestri:2013ne,Lombriser:2014ira} for details. 

Consider a system with some parametrisation of modified gravity and a dust fluid (Cold Dark Matter). 
The continuity and momentum equations of the dust fluid in the quasi-static regime (neglecting the time derivatives of the gravitational potentials), can be condensed into,
\begin{align}
	\frac{d^2\Delta_c(\tau,\vec x)}{d\tau^2}+\mathcal{H}\frac{d\Delta_c(\tau,\vec x)}{d\tau}=-\nabla^2_x\Psi(\tau,\vec x)\;,\label{contineq}
\end{align}
which can be compared to Eqs.~(\ref{eq:energyConservation}, \ref{eq:velocityEquation}) by setting $w=c_s^2=q=\Delta_P=B=\sigma=0$,  $A=\Psi$ and $\dot \Phi = 0$, where the last equalities follows from the quasi-static approximation.
Since, for simplicity, we only consider the dust and gravity sectors, the anisotropic stress term from relativistic components, such as massive neutrinos, {\em etc.} vanishes. 

Combining equation (\ref{modfdPoisson}) and (\ref{gamma}), 
and keeping only terms at linear order,
we get 
\begin{align}
 \nabla_x^2\Psi\simeq -4\pi Ga^2Q(\tau,\vec x)R(\tau,\vec x)\bar\rho_c\Delta_c(\tau,\vec x)\;.\label{modfdPoisson2}
\end{align}
Inserting equation (\ref{modfdPoisson2}) into (\ref{contineq}), we arrive at the equation governing the growth of the dust fluid perturbations~\cite{Dossett:2014oia},
\begin{align}
  \frac{d^2\Delta_c(\tau,\vec x)}{d\tau^2}+\mathcal{H}\frac{d\Delta_c(\tau,\vec x)}{d\tau} - 4\pi G a^2Q(\tau,\vec x)R(\tau,\vec x)\bar\rho_c\Delta_c(\tau,\vec x) =0 \;.\label{mastereq}
\end{align}
In General Relativity, $Q(\tau,\vec x)$ and $R(\tau,\vec x)$ are unity, hence the perturbations of the dust fluid grow in the same way on all the scales, {\em i.e.} the growth rate is only a function of time. However, this is no longer true in the case of modified gravity.

The next step is to calculate the growth rate $\mathcal D$ of CDM density perturbations. In GR, in the late-time and small scale, the growth rate only depends on time; while in the modified gravity case, $\mathcal D(\tau,\vec x)$ depends both on time and space, {\em i.e.}, 
\begin{align}
\Delta_c(\tau,\vec x)=D(\tau,\vec x)\Delta_c(\tau_i,\vec x)\;,\label{growthrate}
\end{align} 
where $\Delta_c(\tau_i,\vec x)$ denotes for the CDM density perturbation at initial time $\tau_i$. 
In order to set initial conditions for a Newtonian N-body simulation, one now uses $\Delta_c$ for the particle positions. Moreover, by `virtue' of the quasi-static approximation, the time derivatives of the potentials are ignored, such that bare quantities are equal to the full physical quantities. In other words, both Newtonian and relativistic (longitudinal gauge) particle positions are obtained from,
\begin{align}
 \vec x=\vec y-\mathcal{D}(\tau,\vec y)\frac{1}{\nabla_y^2}\vec\nabla_y\Delta_c(\tau_i,\vec y)-\Delta_c(\tau_i,\vec y)\frac{1}{\nabla_y^2}\vec\nabla_y\mathcal D(\tau,\vec y)\;.\label{VHA}
\end{align}
Equation (\ref{VHA}) is the main result of this section. If we assume that $\mathcal D$ depends on time only, we recover the Zel'dovich Approximation, which tells us that in General Relativity, at the linear scale, the dust particles propagate on a straight line, because $\vec v_c=\vec\nabla_x\Delta_c(\vec x)$; however, in the modified gravity scenario, the dust particles are also deflected by the gravitational potential due to the second term of equation (\ref{VHA}). 

Finally, in Figure~\ref{fig:curves} we show a discrete realisation of Dark Matter particle trajectories for a particular choice of EFT parameters (see the following section), without the quasi-static approximation. The particle trajectories are compared their $\Lambda$CDM counter parts, and clearly show a curved shape. This figure shows how particle trajectories at the linear level are deflected by modifications of gravity.

\section{Effective Field Theory parametrisation, and modified gravity as a particle displacement field\label{sec:demgeftLPT}}
\subsection{Transfer functions}
After having discussed the physical picture, now let us go through a more advanced parametrisation method of DE/MG models. In this section, we will focus on the effective field theory (EFT) treatment of the cosmic acceleration. This approach is introduced into the study of DE/MG models by \cite{Gubitosi:2012hu,Bloomfield:2012ff} to unify the most generic viable single scalar field models of DE/MG. Let us briefly summarise it here.

Compared with the covariant approach, the construction of the effective action of the scalar field starts with a particular choice of time coordinate, in which the scalar field perturbations vanish. In other words, we first break the four-dimensional covariant diffeomorphisms by choosing a particular clock. Then, we can build all the operators that are consistent with the unbroken symmetries of the theory, {\it i.e.} time-dependent spatial diffeomorphisms. Thanks to this symmetry guidance, we can figure out the spatial structure of these operators. Hence, we are left with only time-dependent prefactors of these operators, namely EFT functions, which need to be parametrized. Another advantage of this procedure is that at the linear order there exist only a few relevant operators which could cover most of the generic viable single scalar field DE/MG models. For these reasons, the EFT approach makes analysing DE/MG models with the on-going and up-coming cosmological surveys feasible.  

The EFT approach relies on the assumption  of the validity of  the weak equivalence principle which ensures the existence of a metric universally coupled to matter fields and therefore of a well-defined Jordan frame. The EFT action in conformal time reads

\begin{align}\label{full_action_Stuck}
S = \int d^4x& \sqrt{-g} \left \{ \frac{m_0^2}{2} \l[1+\Omega(\tau+\pi)\r]R+ \Lambda(\tau+\pi) - c(\tau+\pi)a^2\left[ \delta g^{00}-2\frac{\dot{\pi}}{a^2} +2\hub\pi\left(\delta g^{00}-\frac{1}{a^2}-2\frac{\dot{\pi}}{a^2}\right)\right.\right. \nonumber\\
&\left.\left. +2\dot{\pi}\delta g^{00} +2g^{0i}\partial_i\pi-\frac{\dot{\pi}^2}{a^2}+ g^{ij}\partial_i \pi \partial_j\pi -\l(2\hub^2+\dot{\hub}\r)\frac{\pi^2}{a^2}  \right]+\cdots\right\} + S_{m} [g_{\mu \nu},\chi_i]\;,
\end{align}
where $m_0^2$ is the Planck mass, and $S_m$ is the action for all matter fields, $\chi_i$. For simplicity, here we only list the three operators $\{\Omega$,$\Lambda$,$c\}$ which is describing the background dynamics. For the complete description of the quadratic EFT action, we refer to \cite{Hu:2013twa,Gubitosi:2012hu,Bloomfield:2012ff}.  In the following calculation, we follow the convention of \cite{Hu:2013twa}. 

\begin{figure}[t!]
\begin{center}
  \includegraphics[width=0.45\textwidth]{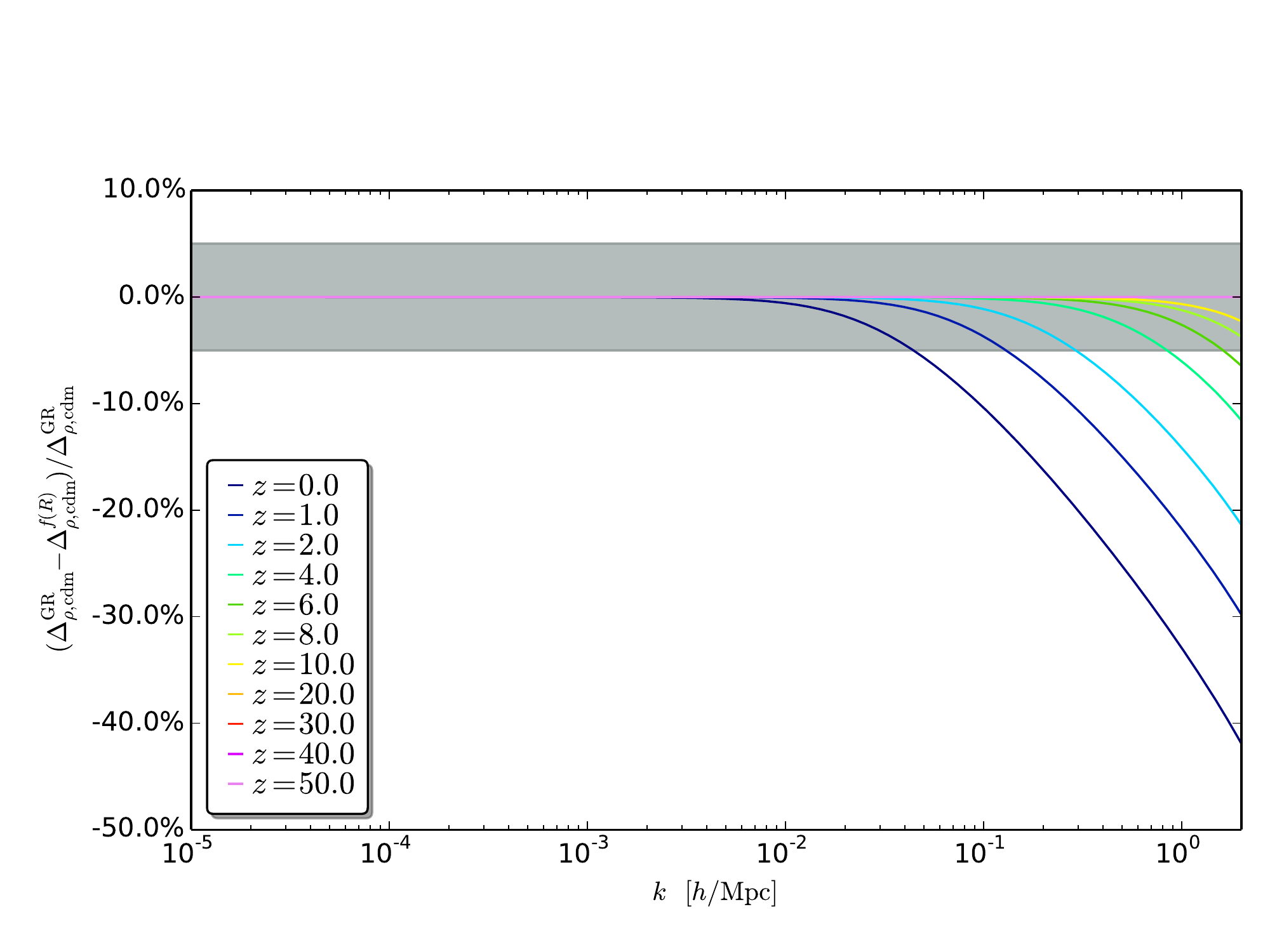}
  \includegraphics[width=0.45\textwidth]{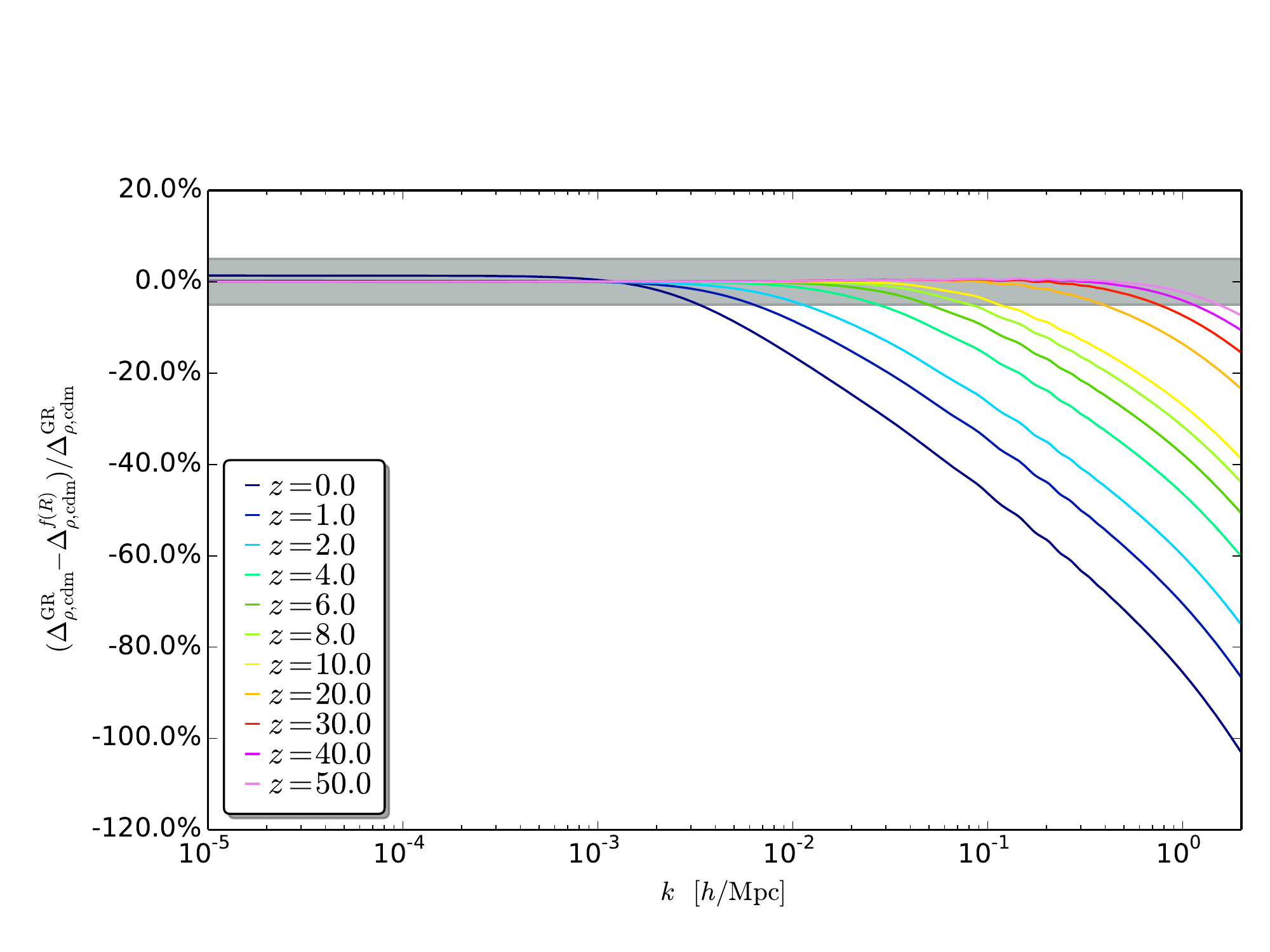}
  \caption{\label{fig:GR_MG_transfer} The fractional differences of the CDM density transfer function between GR and the viable $f(R)$ gravity for several redshift snapshots. (Left panel) Designer $f(R)$ gravity in $\Lambda$CDM background with $B_0=0.001$ and (Right panel) Designer $f(R)$ gravity in wCDM background ($w=-0.95$) with $B_0=0.01$. The grey band denotes for the $5\%$ regime.}
\end{center}
\end{figure}

\begin{figure}[t!]
\begin{center}
  \includegraphics[width=0.45\textwidth]{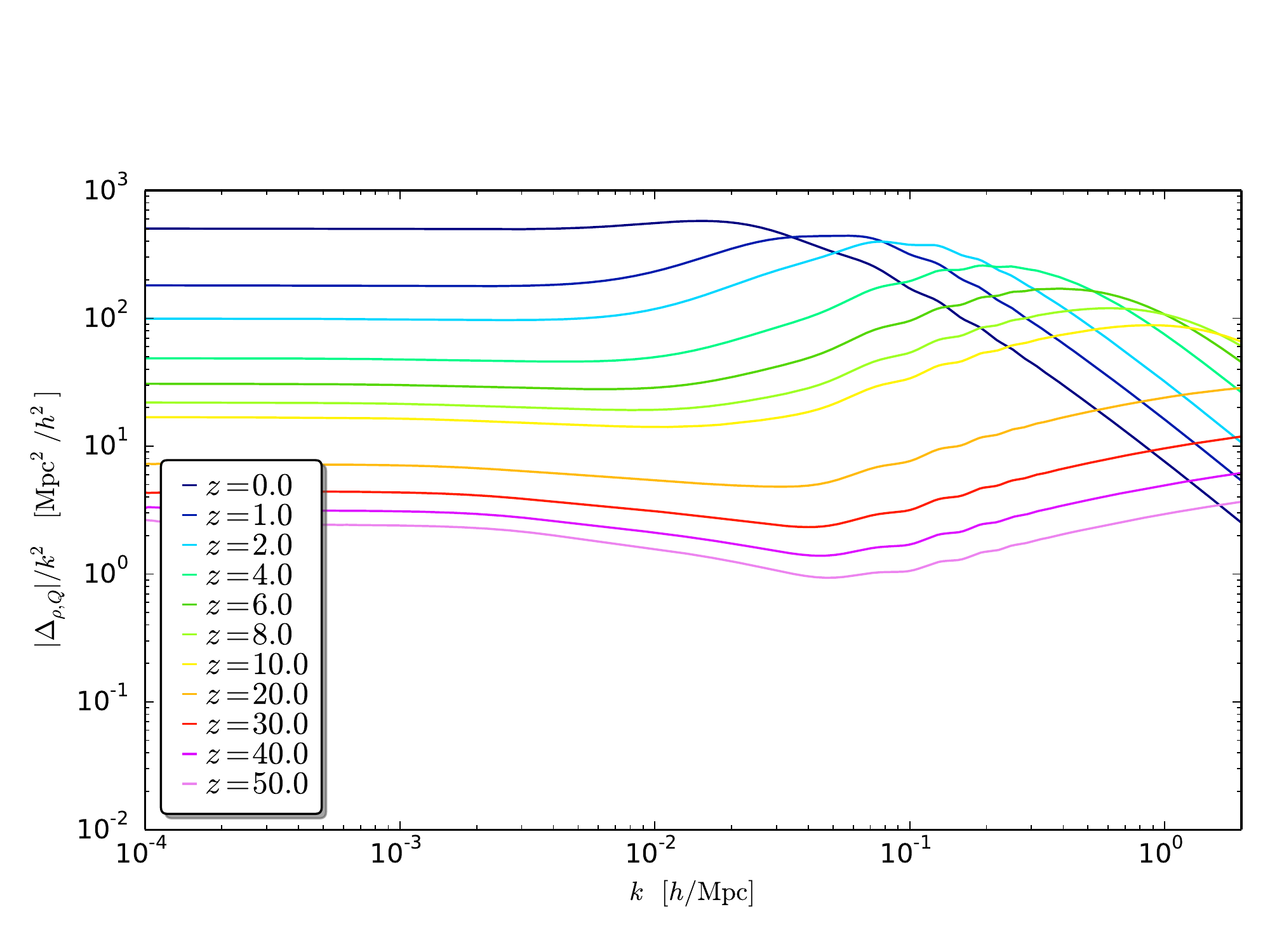}
  \includegraphics[width=0.45\textwidth]{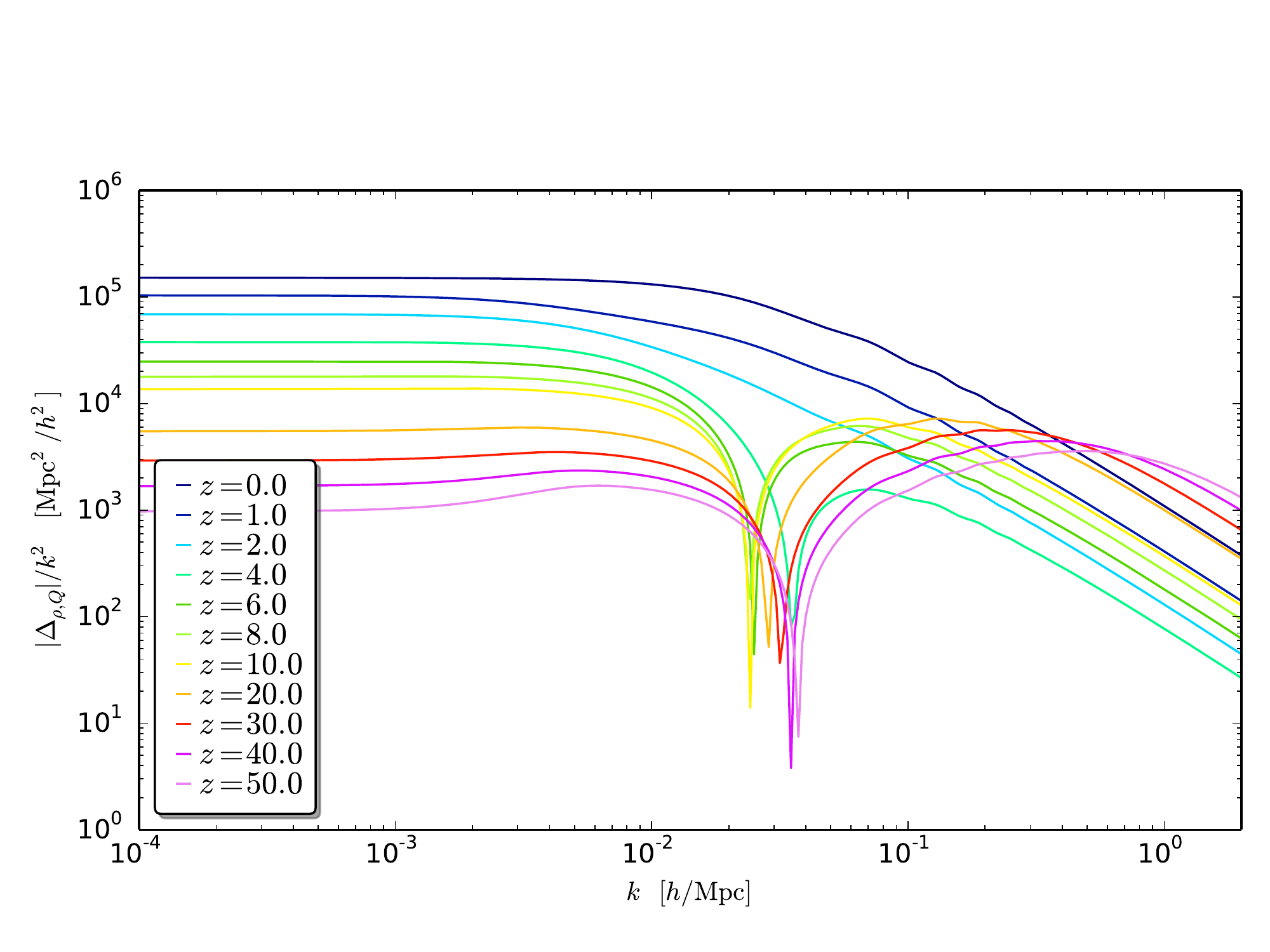}
  \caption{\label{fig:transferq} The $Q$-fluid energy density fluctuation transfer function in two viable $f(R)$ gravity. (Left panel) Designer $f(R)$ gravity in $\Lambda$CDM background with $B_0=0.001$ and (Right panel) Designer $f(R)$ gravity in $w$CDM background ($w=-0.95$) with $B_0=0.01$.}
\end{center}
\end{figure}

\begin{figure}[t!]
\begin{center}
  \includegraphics[width=0.85\textwidth]{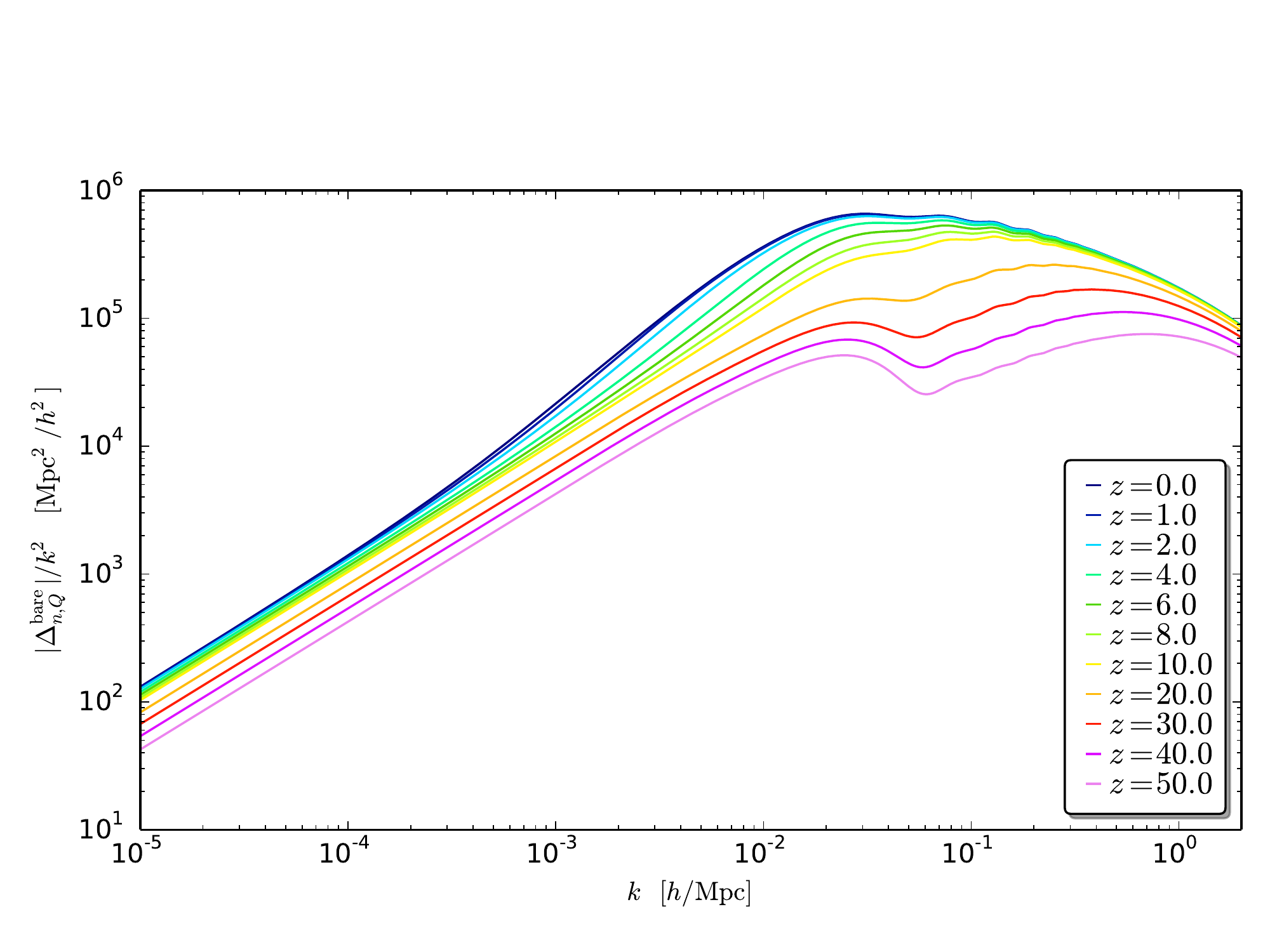}
  \caption{\label{fig:transfer_dnq} The $Q$-fluid bare number density fluctuation transfer function of the designer $f(R)$ gravity in $w$CDM background ($w=-0.95$) with $B_0=0.01$.}
\end{center}
\end{figure}

The modified Einstein equation can be written as
\begin{align}
\label{eq:fluid}
m_0^2(1+\Omega) G_{\mu\nu}[g_{\mu\nu}]=T^{(m)}_{\mu\nu}[\rho_m,\theta_m,\cdots]+ T^{(\pi)}_{\mu\nu}[\pi,\dot\pi,\cdots]\;,
\end{align}
We can divide both sides of the above equation by a factor ($1+\Omega$) and define an effective energy momentum tensor ($T^{(Q)}_{\mu\nu}$), which is conserved by construction ($\nabla^{\nu}T^{(Q)}_{\mu\nu}=0$)\footnote{There exists another equivalent definition by moving the non-minimally coupling term from the left-hand side to the right, {\it i.e.} $T^{(Q)}_{\mu\nu}[\rho_{\pi},\theta_{\pi},g_{\mu\nu},\cdots]\equiv T^{(\pi)}_{\mu\nu}[\pi,\dot\pi,\cdots] - m_0^2\Omega G_{\mu\nu}[g_{\mu\nu}]$. The differences between this definition and the one in equation (\ref{eq:def_QT}) is that in the former, the extra argument of $T^{(Q)}_{\mu\nu}$ is a metric field, while in the latter it is a matter field.}
\begin{align}
m_0^2 G_{\mu\nu}[g_{\mu\nu}]&=T^{(m)}_{\mu\nu}[\rho_m,\theta_m,\cdots]+ T^{(Q)}_{\mu\nu}[\rho_{\pi},\theta_{\pi},\rho_m,\cdots]\;\;, \label{eq:def_QT0}\\
T^{(Q)}_{\mu\nu}[\rho_{\pi},\theta_{\pi},\rho_m,\cdots]&\equiv \frac{1}{1+\Omega}\Big\{-\Omega T^{(m)}_{\mu\nu}[\rho_m,\theta_m,\cdots] +T^{(\pi)}_{\mu\nu}[\pi,\dot\pi,\cdots]\Big\}\;.\label{eq:def_QT}
\end{align}
The reason why we introduce this effective energy momentum tensor ($T^{(Q)}_{\mu\nu}$) instead of directly solve the modified Einstein equation is that we want to apply the Lagrangian perturbation treatment for this effective DE/MG fluid, hereafter $Q$-fluid. The motivation of our algorithms are mainly the following two. 

First of all, in an N-body simulation, a high resolution of the extra scalar field in Eulerian representation is quite expensive, especially for the collapsing DE/MG models.  %
 {This is because in the DE/MG simulation, whether we model the extra scalar degree of freedom as the particle or fluid, depends on the comparison of the mean free path of the particles with the scales we are interested in. Take cold dark matter and massive neutrino as examples, on the scales which are much larger than their mean free path, we can adopt the fluid approximation (ideal fluid for CDM; imperfect fluid for massive neutrino). However, once we concern on the scales which are comparable or even smaller than their mean free path, we have to use the particle description to study its non-linear dynamics. A similar case happens to the collapsing DE/MG model, which has a small sound speed (large mean free path). } %
One solution is to discretise the fluid element into virtual particles with a charge, {\it e.g.} mass, and let the grid follow these virtual particles  {($Q$-particles)}, {\it i.e.} the Lagrangian perturbation approach. This gives the simulation a high resolution at the high density regions.

Second of all, the algorithms to add extra fluid components into N-body and hydro simulation are being extensively developed. We can utilise these techniques to develop the DE/MG simulations via this $Q$-fluid description, such as loading pressure, {\it etc.} In this work, we focus on the linear phenomena, such as linear structure growth and the IC for N-body simulation. In principle, this Lagrangian treatment can be extended to the non-linear phenomena in the simulations.  {Basically the recipe is to generate the initial conditions for the $Q$-particles from the linear Boltzmann code; and then, evolve them with the geodesics.}

Let us go back to the linear perturbation description. Within this approach the above equations (\ref{eq:def_QT0}) and (\ref{eq:def_QT}) can be split into two sets of equations, namely background and (linear) perturbation. On the background, due to historical reason, $\rho_Q$ and $P_Q$ are not the conserved background quantities in the non-minimally coupled case, 
\begin{align}
\rho_Q &= 2c-\Lambda-\frac{3m_0^2\mathcal H\dot\Omega}{a^2}\;,\\
P_Q &=\Lambda+\frac{m_0^2}{a^2}(\ddot\Omega+\mathcal H\dot\Omega)\;.
\end{align}
The conserved one are defined as 
\begin{align}
\rho_{\rm DE} &=-\frac{\Omega}{1+\Omega}\rho_m+\frac{\rho_Q}{1+\Omega}\;,\\
P_{\rm DE} &= -\frac{\Omega}{1+\Omega}P_m+\frac{P_Q}{1+\Omega}\;.
\end{align}

At the linear perturbation level, equation (\ref{eq:def_QT}) and (\ref{eq:def_QT0}) reads
\begin{align}
\label{eq:fluid}
m_0^2\delta G_{\mu\nu}[\delta g_{\mu\nu}]&=\delta T^{(m)}_{\mu\nu}[\delta\rho_m,\theta_m,\cdots]+\delta T^{(Q)}_{\mu\nu}[\delta\rho_{\pi},\theta_{\pi},\delta\rho_m,\cdots]\;,\\
\delta T^{(Q)}_{\mu\nu}[\delta\rho_{\pi},\theta_{\pi},\delta\rho_m,\cdots]&=\frac{1}{1+\Omega}\Big\{-\Omega T^{(m)}_{\mu\nu}[\delta\rho_m,\theta_m,\cdots]+\delta T^{(\pi)}_{\mu\nu}[\pi,\dot\pi,\cdots]\Big\}\;.
\end{align}
Armed with these results, we could recognize the fluid variables, such as the energy density, velocity, pressure as well as anisotropic stress tensor. Within the fully relativistic treatment, the definition of these quantities are gauge related. In the synchronous gauge, the Einstein equation reads
\begin{align}
-\frac{2m_0^2}{a^2}\Big(k^2\eta-\frac{1}{2}\mathcal H\dot h\Big)&=\delta\rho^{({\rm syn})}_m+\delta\rho^{({\rm syn})}_Q\;,\\
\frac{2m_0^2}{a^2}k^2\dot\eta&=(\rho_m+P_m)\theta_m^{({\rm syn})}+(\rho_{\rm DE}+P_{\rm DE})\theta_Q^{({\rm syn})}\;,\\
-\frac{m_0^2}{a^2}\Big(\ddot h+2\mathcal H\dot h-2k^2\eta\Big)&=3\delta P_m^{({\rm syn})}+3\delta P_Q^{({\rm syn})}\;,\\
-\frac{m_0^2}{3a^2}\Big[\ddot h+6\ddot\eta+2\mathcal H(\dot h+6\dot\eta)-2k^2\eta\Big]&=(\rho_m+P_m)\sigma_m^{({\rm syn})}+(\rho_{\rm DE}+P_{\rm DE})\sigma_{Q}^{({\rm syn})}\;,
\end{align}
where the fluid variables are defined as
\begin{align}
\delta\rho_{Q}^{({\rm syn})}&=\frac{1}{(1+\Omega)}\left\{-\Omega\delta\rho_m^{({\rm syn})}+\dot\rho_{Q}\pi+2c(\dot\pi^{({\rm syn})}+\mathcal H\pi^{({\rm syn})})\right.\nonumber\\
&\left.-\frac{2m_0^2}{a^2}\left[\frac{\dot\Omega}{4}\dot h+\frac{\dot\Omega}{2}\Big(3(3\mathcal H^2-\dot{\mathcal H})\pi^{({\rm syn})}+3\mathcal H\dot\pi^{({\rm syn})}+k^2\pi^{({\rm syn})}\Big)\right]\right\}\;,\label{eq:Qdensity}\\
(\rho_{\rm DE}+P_{\rm DE})\theta_Q^{({\rm syn})}&=\frac{1}{1+\Omega}\left[-\Omega(\rho_m+P_{m})\theta_m^{({\rm syn})}+(\rho_Q+P_Q)k^2\pi^{({\rm syn})}\right.\nonumber\\
&+\left.\frac{2m_0^2}{a^2}k^2\dot\Omega(\dot\pi^{({\rm syn})}+\mathcal H\pi^{({\rm syn})})\right]\;,\label{eq:thetaq}\\
\delta P_Q^{({\rm syn})}&=\frac{1}{1+\Omega}\left\{-\Omega\delta P_m^{({\rm syn})}+P_Q\dot\pi^{({\rm syn})}+(\rho_Q+P_Q)(\dot\pi^{({\rm syn})}+\mathcal H\pi^{({\rm syn})})\right.\nonumber\\
&+\left.\frac{m_0^2}{a^2}\left[\frac{1}{3}\dot\Omega\dot h+\dot\Omega\ddot\pi^{({\rm syn})}+(\ddot\Omega+3\mathcal H\dot\Omega)\dot\pi^{({\rm syn})}+\left(\mathcal H\ddot\Omega+5\mathcal H^2\dot\Omega+\dot{\mathcal H}\dot\Omega+\frac{2}{3}k^2\dot\Omega\right)\pi^{({\rm syn})}\right]\right\}\;,\\
(\rho_{\rm DE}+P_{\rm DE})\sigma_{Q}^{({\rm syn})}&=\frac{1}{1+\Omega}\left[-\Omega(\rho_m+P_m)\sigma_m^{({\rm syn})}+\frac{m_0^2}{3a^2}\dot\Omega\Big(\dot h+6\dot\eta+2k^2\pi^{({\rm syn})}\Big)\right]\;. 
\end{align}
Beware that in the $\Lambda$CDM background ($\rho_{\rm DE}+P_{\rm DE}=0$), the divergence of the velocity $\theta_Q^{({\rm syn})}$ and the anisotropic stress tensor $\sigma_{Q}^{({\rm syn})}$ of the $Q$-fluid are not well defined. The Einstein equation and the $Q$-fluid energy momentum tensor in the longitudinal gauge are listed in the appendix \ref{App1}. 

In the rest of this section, we will evolve the full dynamics of the $Q$-fluid (or $\pi$ field) by using {\sc EFTCamb} \cite{Hu:2013twa,Raveri:2014cka}, calculate the relevant fluid quantities for the simulations and produce a realisation of discrete initial conditions for a discrete simulation. In the following numerical calculations, for simplicity, we take the $f(R)$ gravity as an example to show the non-trivial dynamics of the $Q$-fluid. 
In details, we investigate the designer $f(R)$ model \cite{Song:2006ej,Pogosian:2007sw} in $\Lambda$CDM and $w$CDM background ($w=P_{\rm DE}/\rho_{\rm DE}$), which could {\it exactly} reproduce the background history we fixed {\it a priori}. After fixing the background expansion history, the extra degree of freedom of this higher derivative gravity theory still allows for one extra parameter, here we choose it as the present value of the Compton wavelength of the scalar field, namely $B_0\sim \frac{6 f_{RR}}{(1+f_{R})}H^2|_{a=1}$ (in Hubble parameter unit) \cite{Hu:2007nk}, where $f_R\equiv \partial_R f$. We take the $B_0$ parameter value inside the {\it viable} regime after Planck-2013 results \cite{Raveri:2014cka}, namely for designer $f(R)$ model in $\Lambda$CDM case we take $B_0=0.001$; for designer $f(R)$ model in $w$CDM case we take ($B_0=0.01,w=-0.95$). 

In Figure~\ref{fig:GR_MG_transfer} we show the fractional differences between GR and the designer $f(R)$ models, for the cold dark matter density perturbations  at 11 redshift snapshots. The left panel for the $\Lambda$CDM-mimicking case with $B_0=0.001$ ($w=-1$) and the right panel for the $w$CDM-mimicking case with ($B_0=0.01, w=-0.95$). The grey band in the figures is the $5\%$-deviation regime. We can see that for the $\Lambda$CDM case at redshift $z=50$ (pink curve), where generally N-body simulations start to run, the differences are deep inside the $5\%$ regime. However, for the $w$CDM case on scales smaller than ($k\gtrsim 1~[h/{\rm Mpc}]$) the differences are outside of the grey band. 

In Figure~\ref{fig:transferq} we show the transfer functions of the energy density of the $Q$-fluid. Our calculation demonstrates that for the $\Lambda$CDM case, $\delta\rho_Q$ is anti-correlated with $\delta\rho_{\rm cdm}$ on all the scales, {\it i.e.} in the CDM over-dense region ($\delta\rho_{\rm cdm}>0$) the $Q$-fluid energy density perturbation keeps under-dense ($\delta\rho_Q<0$).  This reflects the fact that the CDM perturbation in the $f(R)$ gravity is enhanced in the linear regime. To explain this, we take the Poisson equation in the Newtonian limit
\begin{align}
-k^2\psi&=\frac{16\pi G}{3}\delta\rho_{\rm cdm}-\frac{\delta R}{6}\;.\label{eq:for}
\end{align}
In the CDM over-dense region, $\delta\rho_{\rm cdm}>0$, the Newtonian potential $\psi<0$ and $\delta R>0$. In our $Q$-fluid language, equation (\ref{eq:for}) reads
\begin{align}
-k^2\psi&=\frac{16\pi G}{3}(\delta\rho_{\rm cdm}+\delta\rho_Q)\;.\label{eq:for2}
\end{align}
Comparing these two equations, we get $\delta\rho_Q\propto -\delta R$ and this quantity stays negative. In order to generate the same depth of the gravitational potential well ($\psi$), we need extra CDM fluctuations to compensate the negative $\delta\rho_Q$. This indicates the fact that the growth rate of CDM get enhanced. From the modified gravity point of view at the linear scale\footnote{Here the ``linear scale'', we mean the scale above the screening scale via the chameleon mechanism \cite{Brax:2004qh}.}, the effective gravitational constant $G_{\rm eff}$ is enhanced by a factor $4/3$ compared with those in GR. What we find is nothing new, just another explanation of the same phenomena in term of an exotic fluid component.  From the right panel of Figure~\ref{fig:transferq} we can see that in the deep redshift, the $Q$-fluid energy density perturbation change the sign (on the large scale $\Delta_{\rho,Q}$ is positive; while on the small scale $\Delta_{\rho,Q}$ is negative\footnote{Our the above explanation is still valid in the $w$CDM case on the small scale, where the collapsing of Dark Matter will happen.}), but in the late redshift it does not. This is due to the complicated competition relationship between the $\delta\rho_m$ and $\pi$ field in the definition of $\delta\rho_Q$ in equation (\ref{eq:Qdensity}).

In Figure~\ref{fig:transfer_dnq} we show the transfer function of the bare number density perturbation (\ref{eq:dnbare}) of the $Q$-fluid ($\Delta_{n,Q}^{\bare}$). As we have discussed in section (\ref{sec:bare}), this quantity defines the positions of vertices at rest with the flow of the $Q$-fluid, in relativistic simulations. From the equation (\ref{eq:baredot}) we can see that $\Delta_{n,Q}^{\bare}$ is the integration of $\theta_Q$ over the conformal time. Furthermore, from equation (\ref{eq:thetaq}) we know that in the $\Lambda$CDM background $\theta_Q$ is not well defined, neither is $\Delta_{n,Q}^{\bare}$. Given these considerations, in Figure~\ref{fig:transfer_dnq} we only show $\Delta_{n,Q}^{\bare}$ in the $w$CDM background case. We can see that, unlike $\Delta_{\rho,{\rm cdm}}/k^2$, the quantity $\Delta_{n,Q}^{\bare}/k^2$ increases on small scales. This also reflects the fact that the modification of gravity in $f(R)$ models becomes more significant on smaller scales,  though above the screening scale. 

\subsection{Initial conditions for discrete simulations}
\begin{figure}
\begin{center}
\begin{tabular}{ccc}
&{\em Relativistic and Eulerian}&\\
\includegraphics[width=0.3\textwidth]{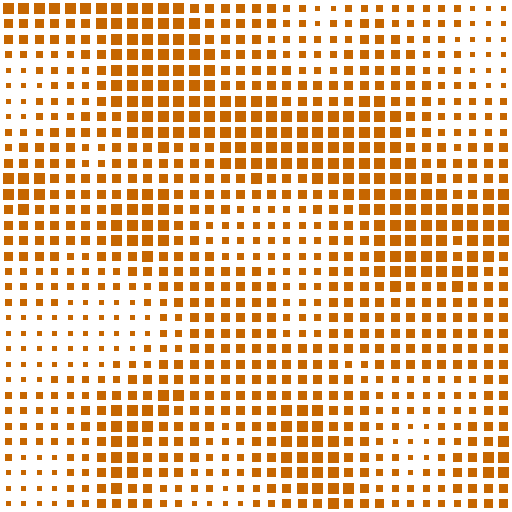}&
\includegraphics[width=0.3\textwidth]{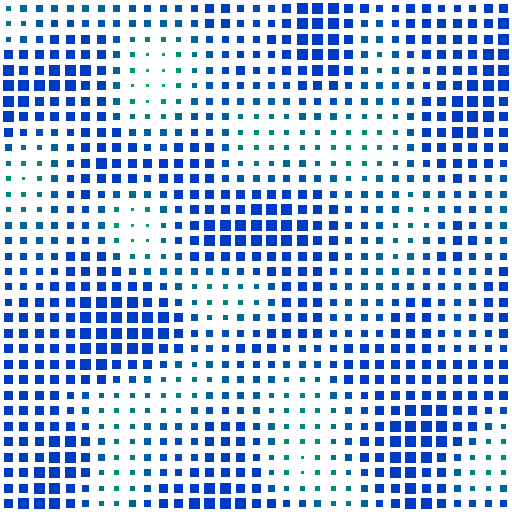}&
\includegraphics[width=0.3\textwidth]{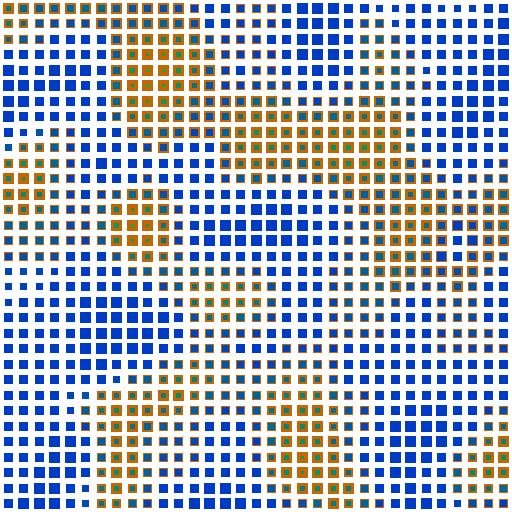}\\
Dark Matter (Eulerian) & $Q$-fluid (Eulerian) & Combined\\
&&\\&&\\
&{\em Newtonian and Lagrangian}&\\
\includegraphics[width=0.3\textwidth]{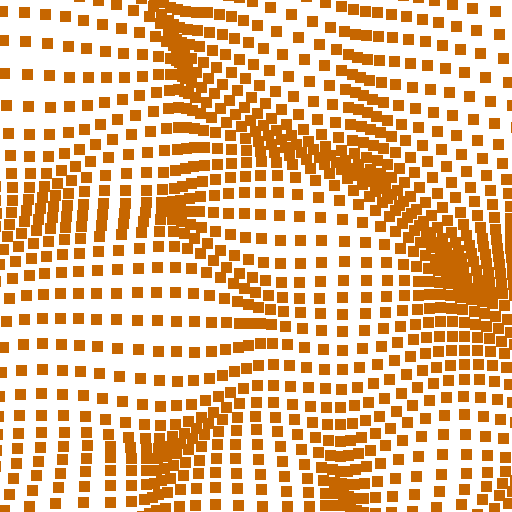}&
\includegraphics[width=0.3\textwidth]{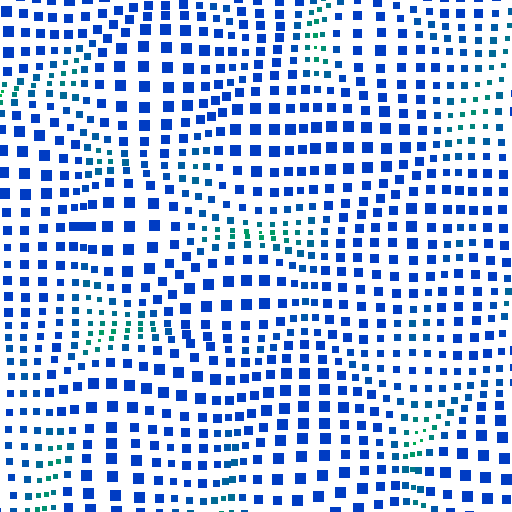}&
\includegraphics[width=0.3\textwidth]{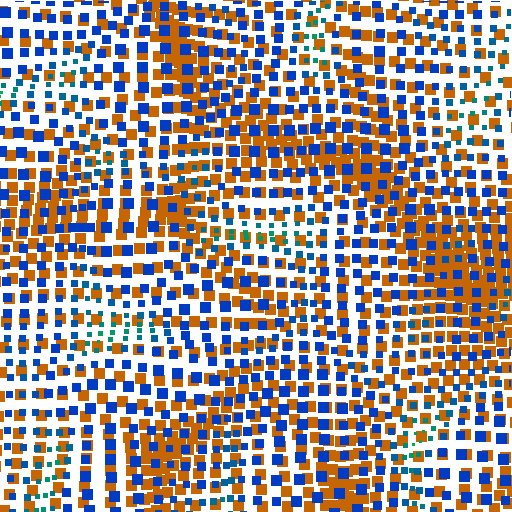}\\
Dark Matter  & $Q$-particles (Langrangian) & Combined\\
(Langrangian, exaggerated)&&
\end{tabular}
\end{center}
\caption{
Slices of realizations of initial conditions at $z=50$, for use in relativistic (top row) and newtonian (bottom row) simulations, based on linear perturbation solutions from {\sc EFTCamb}. The setup is identical to the $f(R)$-cosmology in Figure~\ref{fig:curves}. The space between vertices is $1.56$ Mpc. Sizes of the squares represent masses of the particles (densities at the vertices), while for the $Q$-fluid, the color additionally represents internal energy (pressure at the vertices), where the colour green indicates a low pressure, which simply traces under-densities.  We omit the combination `Relativistic and Lagrangian', since {\sc camb} works in synchronous gauge, and the velocity of Dark Matter in that gauge is zero, and a Lagrangian representation in that gauge is meaningless. The $Q$-fluid can be treated in the Lagrangian Representation in a relativistic simulation in coordinates synchronous and comoving with Dark Matter. A relativistic simulation of Dark Matter only makes sense in a gauge that does not produces caustics, such as for example the longitudinal gauge. The top row serves as a proof of concept. The displacement of Dark Matter is exaggerated by a factor of 20 in the bottom row, for the purpose of contrast in the illustration.
}
\label{fig:examplesEL}
\end{figure}

In Figure~\ref{fig:examplesEL} we show an application of all the findings in this paper. {  We discretise the results of the previous sections using a discretisation as described in Appendix~\ref{sec:discrete}.}  In the top row, we setup initial conditions for a relativistic simulation in synchronous comoving coordinates, as a proof of concept. Such a simulation would make little sense, as the motivation for a simulation with Dark Matter is to be able to trace the particles beyond shell crossing, which invalidates the synchronous gauge.  The figures show a regular grid, which is comoving with the Dark Matter. Proper  masses (not bare!) are indicated by the size of squares. Moreover, for the $Q$-fluid a low pressure is indicate the colour green, while high pressure is indicated in blue. For this fluid, pressure simply traces density.

In the bottom row of~\ref{fig:examplesEL}, we show the Newtonian approximation based on the same spectra computed in the synchronous comoving gauge, as explained in Section~\ref{sec:nrlim}. In this case, both fluids can be considered in the Lagrangian representation, that is, represented by vertices (particles) which are at rest with each species' flow. From this picture it becomes explicit that the phrase `at rest with the flow' does not imply that mass- and internal-energy densities are constant in such a frame, if pressure or heat transfer are present. For Dark Matter, pressure is zero and hence these particles do have constant mass (equal-sized squares) in the Lagrangian representation\footnote{If there is a direct coupling between Dark Matter and the extra scalar field, the Dark Matter might not be conserved due to the non-trivial coupling\cite{Amendola:1999er,Matarrese:2003tn,Maccio:2003yk,Perrotta:2003rh}.}.

In the Eulerian view, it is clearly seen that the $Q$-fluid has density perturbations which are anti-correlated with the Dark Matter perturbations, since both fluids are over-dense in complementary regions. This exemplifies the discussion at Eqs.~(\ref{eq:for},~\ref{eq:for2}). However, it is important to realise that this regular grid is in the frame comoving with the Dark Matter, which is hence not a regular grid in newtonian coordinates; in the newtonian representation (bottom row), one can recognise that the low-mass and low-pressure vertices in the $Q$-fluid are in fact clustered, partially compensating the density perturbation that is visible in the top row.

\section{Conclusion}

We have provided the means to generate a distribution of particles with masses, positions and internal energies such that they describe exactly the underlying fluid theory, for both perfect and imperfect fluids, in arbitrary gauges. That is, simulations in longitudinal, synchronous co-moving, or the so-called `N-body gauge'~\cite{Fidler:2015npa}, find their initial conditions following our prescription. The newtonian limit is always found by taking the dressed densities for particle positions, while fully relativistic particle positions are defined by the bare number densities. The description holds for any metric theory of gravity, including the newtonian limit of General Relativity. The description can be used in any gauge, for any description of a fluid, with velocities defined as comoving with any quantity of the fluid.

We applied our findings to a description of Modified Gravity, in which we describe the extra degree of freedom as an imperfect fluid $Q$, with particles with varying mass and internal energy. Trajectories of all species in the linear regime are curves rather than straight lines. Most importantly, we show that the perturbations in new dynamical degrees of freedom cannot be ignored when setting initial conditions. 

{  Such initial conditions can be applied in at least three ways: (1) when the modified gravity field stays linear and its fluid description remains valid, a Lagrangian representation of the linear field can be included in an N-body simulation during all times, providing high resolution where the simulation needs it, (2) when the nonlinear dynamics of the field are still described by a fluid, the field can be simulated by adapting existing hydrodynamics techniques, and (3) these initial conditions can be used for any particle species whose linear solution is effectively described by an imperfect fluid, such as neutrinos on scales larger than their free-streaming length.}

Our findings open the way to modelling non-Newtonian gravity in various ways, in arbitrary coordinate choices. The advantages of this possibility remain to be explored. 

We release a C$^{++}$-code for the generation of initial conditions, FalconIC, at {\url{http://falconb.org}}, with a minimalistic GUI that runs natively on Linux and OS X. The code links against any version of both Boltzmann codes {\sc camb} and {\sc class}, including {\sc EFTCamb}, such that no separate running of those codes is necessary, and generates initial conditions at arbitrary scales, of arbitrary size (fully parallelised using MPI and OpenMP), for arbitrary cosmological parameters.

\section*{Acknowledgements}
The authors wish to thank 
Ignacy Sawicki,
Julien Bel, 
Carmelita Carbone,
Julien Lesgourgues,
Mark Lovell,
Ewald Puchwein,
Cornelius Rampf, 
Marco Raveri,
Gerasimos Rigopoulos,
Christoph Schmid
and
Alessandra Silvestri
for fruitful discussions. 
WV is supported by a Veni research grant from the Netherlands Organization for Scientific Research (NWO).
BH is supported by the Dutch Foundation for Fundamental Research on Matter (FOM). 

\appendix
\section{Gauge transformation and $Q$-fluid variables in the longitudinal gauge\label{App1}}
In the longitudinal gauge, with the convention of Ma and Bertschinger \cite{Ma:1995ey},
\begin{align}
ds^2&=a^2\Big[-(1+2\psi)d\tau^2+(1-2\phi)\delta_{ij}dx^idx^j\Big]\;,\\
ds^2&=a^2\Big[-(1+2A)d\tau^2+(1+2H_L)\delta_{ij}dx^idx^j\Big]\;,
\end{align}
so we have $A=\psi\;,H_L=-\phi$. In the following, we will move to the ($\psi,\phi$) convention
\begin{align}
\psi&=\frac{1}{2k^2}\Big[\ddot h+6\ddot\eta+\mathcal H(\dot h+6\dot\eta)\Big]\;,\\
\phi&=\eta-\frac{\mathcal H}{2k^2}(\dot h+6\dot\eta)\;,\\
\pi^{\longitudinal}&=\pi^{\synchronous}+\frac{1}{2k^2}(\dot h+6\dot\eta)\;,
\end{align}
In the longitudinal gauge, the Einstein equation reads
\begin{align}
-\frac{2m_0^2}{a^2}\Big[k^2\phi+3\mathcal H(\dot\phi+\mathcal H\psi)\Big] &=\delta\rho_m^{\longitudinal}+\delta\rho_Q^{\longitudinal}\;,\\
\frac{2m_0^2}{a^2}k^2\Big(\dot\phi+\mathcal H\psi\Big) &=(\rho_m+P_m)\theta_m^{\longitudinal}+(\rho_{\rm DE}+P_{\rm DE})\theta_Q^{\longitudinal}\;,\\
\frac{2m_0^2}{a^2}\left[\ddot\phi+\mathcal H(\dot\psi+2\dot\phi)+(\mathcal H^2+2\dot{\mathcal H})\psi+\frac{k^2}{3}(\phi-\psi)\right] &= \delta P_m^{\longitudinal}+\delta P_Q^{\longitudinal}\;,\\
\frac{2m_0^2}{3a^2}k^2(\phi-\psi) &= (\rho_m+P_m)\sigma_{m}^{\longitudinal} + (\rho_{\rm DE}+P_{\rm DE})\sigma_{Q}^{\longitudinal}\;.
\end{align}
And the $Q$-fluid variables are defined as
\begin{align}
\delta\rho_Q^{\longitudinal} &=\frac{1}{1+\Omega}\left\{-\Omega\delta\rho_m^{\longitudinal}+\dot\rho_Q\pi^{\longitudinal}+2c(\dot\pi^{\longitudinal}+\mathcal H\pi^{\longitudinal}-\psi)\right.\nonumber\\
&\left.-\frac{m_0^2}{a^2}\dot\Omega\left[3(2\mathcal H^2-\dot{\mathcal H})\pi^{\longitudinal}-3\mathcal H(\dot\pi^{\longitudinal}+\mathcal H\pi^{\longitudinal}-\psi)+k^2\pi^{\longitudinal}-3(\dot\phi+\mathcal H\psi)\right]\right\}\;,\\
(\rho_{\rm DE}+P_{\rm DE})\theta_Q^{\longitudinal} &=\frac{1}{1+\Omega}\left[-\Omega(\rho_m+P_m)\theta_m^{\longitudinal}+(\rho_Q+P_Q)k^2\pi^{\longitudinal}\right.\nonumber\\
&\left.+\frac{m_0^2}{a^2}\dot\Omega k^2(\dot\pi^{\longitudinal}+\mathcal H\pi^{\longitudinal}-\psi)\right]\;,\\
\delta P_Q^{\longitudinal} &=\frac{1}{1+\Omega}\left\{-\Omega\delta P_m^{\longitudinal} +\dot P_Q\pi^{\longitudinal}+(\rho_Q+P_Q)\Big(\dot\pi^{\longitudinal}+\mathcal H\pi^{\longitudinal}-\psi\Big)\right.\nonumber\\
&+\frac{m_0^2}{a^2}(\ddot\Omega-\mathcal H\dot\Omega)\Big(\dot\pi^{\longitudinal}+\mathcal H\pi^{\longitudinal}-\psi\Big)+\frac{m_0^2}{a^2}\dot\Omega\left[\ddot\pi^{\longitudinal}+\mathcal H\dot\pi^{\longitudinal}+\dot{\mathcal H}\pi^{\longitudinal}\right.\nonumber\\
&\left.\left.-\dot\psi-2\dot\phi+3\mathcal H\Big(\dot\pi^{\longitudinal}+\mathcal H\pi^{\longitudinal}\Big)-5\mathcal H\psi+3\mathcal H^2\pi^{\longitudinal}+\frac{2}{3}k^2\pi^{\longitudinal}\right]\right\}\;,\\
(\rho_{\rm DE}+P_{\rm DE})\sigma_Q^{\longitudinal}&=\frac{1}{1+\Omega}\left[-\Omega(\rho_m+P_m)\sigma_m^{\longitudinal}+\frac{2m_0^2}{3a^2}\dot\Omega k^2\pi^{\longitudinal}\right]\;.
\end{align}

{ 
\section{Discrete sampling of phase space \label{sec:discrete}}
For the sake of completeness, here we briefly summarise the currently known methods for generating a discrete sampling of the otherwise continuous phase space of a fluid. For in-depth discussions, we refer the reader to~\cite{Bertschinger:2001ng,Hahn:2011uy}.

In this section we work {\em top down}, in that we start from a theoretical continuous field on an infinite space, and construct a finite-resolution finite-sized discrete representation that resembles the theoretical field as closely as possible.

Discreteness enters at two ends of the length scale: (1) moving from an infinite box size to a finite box size renders the eigen space of the laplacian discrete (Fourier transforms become sums rather than integrals), which sets the lowest nonzero wave number under consideration; and (2) only finitely many numbers can be treated on a discrete computer, which on one hand limits the highest sampled wave number and on the other hand limits the number of simulated vertices (particles).

A statistically isotropic gaussian random field  $\hat s(\vec x)=\int \frac{d^3k}{(2\pi)^3} e^{i\vec k \cdot \vec x} \hat s_{\vec k}$ in an infinite-sized space can be defined by its Fourier-space correlator,
\begin{align}
\left<\hat s_{\vec k}\hat s_{\vec k'}\right> = (2\pi)^3\delta^3(\vec k - \vec k') P_{\hat s}(k),
\end{align}
such that,
\begin{align}
	\left< \hat s(\vec x) \hat s^*(\vec x') \right>=&\int \frac{d^3k}{(2\pi)^3} e^{-\vec k\cdot \left( \vec x - \vec x' \right)} P_{\hat s}(k),\label{eq:gausscorrel}\\
	\left< \left|\hat s(\vec x)\right|^2  \right> =&\int \frac{d^3k}{(2\pi)^3}  P_{\hat s}(k),
\end{align}
is independent of position $\vec x$. Random variables are denoted by a hat, $\hat s$.

\subsection{Finite length}
The continuous field $\hat s(\vec x)$ in the finite-sized box with length $L$ and periodic boundary conditions (a 3-torus) is related to the continuous field in infinite-space,
\begin{align}
	\hat s(\vec x)=&\int \frac{d^3k}{(2\pi)^3} e^{i\vec k \cdot \vec x} \hat s_{\vec k}\nonumber\\
		=& \lim_{L\rightarrow\infty} \sum_{\vec n=-\infty}^{\infty} L^{-3} e^{i\frac{2 \pi \vec n}{L}\cdot \vec x}  \hat s_{\frac{2 \pi \vec n}{L}},\label{eq:discretelims}
\end{align}
where $\vec n$ is a vector of three integers. Inserting Eq.~\eqref{eq:discretelims} into Eq.\eqref{eq:gausscorrel}, one finds that,
\begin{align}
\left< \hat s_{\frac{2 \pi \vec n}{L}}  \hat s^*_{\frac{2 \pi \vec n'}{L}} \right> = L^{-3} \delta_{\vec n, \vec m} P_{\hat s}\left(\left|\frac{2 \pi \vec n}{L}\right|\right),\label{eq:appTwoPointDiscrete}
\end{align}
where $\delta_{\vec n, \vec m}$ is the Kronecker delta.

\subsection{Finite resolution}
Next, keeping the finite box size $L$ (not taking the infinite limit), the relation between the continuous and discrete functions in the box is set by the highest wave number $\pi M/L$,
\begin{align}
	\hat s(\vec x)
		=& \lim_{M\rightarrow\infty} \sum_{\vec n=-M/2}^{M/2} L^{-3} e^{i\frac{2 \pi \vec n}{L}\cdot \vec x}  \hat s_{\frac{2 \pi \vec n}{L}}.\label{eq:appFourierFiniteResolutionLimit}
\end{align}

It is custom to simply take a finite $M$, which is equivalent to multiplying the Fourier space $\hat s_{\frac{2 \pi \vec n}{L}}$ with a tophat filter. Different filters may be chosen. The real space representation of the tophat filter indeed closely resembles a series of unit valued peaks, equally spaced at a separation $L/M$, such that the finite $M$ leads to a real space $s(\vec x)$ which is convolved with a series of delta functions, {\em i.e.} a discrete sampling. Note however that relation between the tophat filter and a discrete real space sampling is not exact, though it is a good approximation. Formally, a discrete periodic sampling of a function in real space corresponds to a discrete periodic sampling in Fourier space exactly, only when both samples are infinite ($M\rightarrow\infty$, see~\cite{bracewell2000fourier}).

When one wants to generate a coarse grained realisation that represents an averaging of the continuous field, the filter applied inside the sum of Eq.~\eqref{eq:appFourierFiniteResolutionLimit} should be the Fourier transform of the filter used in the real space averagin, such as a gaussian filter, cloud-in-cell, triangular-shaped-cloud, and so on~\cite{Cui:2008fi}.

In summary, the continuous gaussian random field $\hat s(\vec x)$ on an infinite space, can be discretely sampled on a grid with length $L$ and resolution $M$ by generating gaussian random numbers that obey,
\begin{align}
	\left<\hat s(\vec n L/M) \hat s(\vec n' L/M)\right> = \sum_{\vec m = -M/2}^{M/2} L^{-3} e^{i{\frac{2\pi}{M}\vec m \cdot(\vec n - \vec n')}} P_{\hat s}\left(\left|\frac{2 \pi \vec m}{L}\right|\right),
\end{align}
with $\vec n$ a vector of three integers in the range $\left[0,M\right)$ and $\vec m$ a vector with integers in the range $\left[-M/2,M/2\right)$.

\subsection{Generation in Fourier space}
From Eq.~\eqref{eq:appTwoPointDiscrete}, we find that both the real and imaginary parts of $ \hat s_{\frac{2 \pi \vec n}{L}} $ are independent gaussian random variables (see {\em e.g.} \cite{Sirko:2005uz}) with variance $\sigma_{\hat s} = \sqrt{\tfrac{1}{2}L^{-3}  P_{\hat s}\left(\left|\frac{2 \pi \vec n}{L}\right|\right) }$, such that, 
\begin{align}
\left< \left|\hat s_{\frac{2 \pi \vec n}{L}} \right|^2  \right> =& \left< {\rm Re\,} \hat s_{\frac{2 \pi \vec n}{L}}^2  \right> + \left< {\rm Im\,} \hat s_{\frac{2 \pi \vec n}{L}}^2  \right> \nonumber\\ 
=& L^{-3} \delta_{\vec n, \vec m} P_{\hat s}\left(\left|\frac{2 \pi \vec n}{L}\right|\right).
\end{align}
Imposing Hermitian symmetry $\hat s^*_{\frac{2 \pi \vec n}{L}} = \hat s_{-\frac{2 \pi \vec n}{L}}$ guarantees that $\hat s(\vec x)$ be real valued. Finally, one simply has to draw normal gaussian random complex values $\hat \xi$ for each discrete wave number, and multiply by its respective amplitude to realise one random field,
\begin{align}
	\hat s(\vec x_i)
		=& \sum_{\vec n=-M/2}^{M/2}  e^{i\frac{2 \pi \vec n}{L}\cdot \vec x_i}  \left[{\tfrac{1}{2}L^{-3}  P_{\hat s}\left(\left|\frac{2 \pi \vec n}{L}\right|\right) }\right]^{\frac{1}{2}} \hat\xi\left(\frac{2 \pi \vec n}{L}\right),\label{eq:appFourierRealisation}
\end{align}
where $\vec x_i$ represents a position on the discrete grid, with three discrete components in the range $\left[0,M\right)$.

\subsection{Generation in real space}
Following \cite{Bertschinger:2001ng}, the separation of amplitude and normal random variables in Eq.~\eqref{eq:appFourierRealisation} implies that $\hat s(\vec x_i)$ can be written as a convolution of the inverse Fourier transforms of the individual ingredients, the amplitude and the normal random field. A normal random field transforms into a normal random field. Hence, one could just as well generate the normal random field in real space, and perform the convolution with the real space transform,
\begin{align}
	T(\vec x_i) =&  \sum_{\vec n=-M/2}^{M/2}  e^{i\frac{2 \pi \vec n}{L}\cdot \vec x_i}  \left[{\tfrac{1}{2}L^{-3}  P_{\hat s}\left(\left|\frac{2 \pi \vec n}{L}\right|\right) }\right]^{\frac{1}{2}}.
\end{align}

Generating just one grid, this method is equivalent to the Fourier space method. However, now having access to the random numbers in real space, this opens the way to generating higher resolution initial conditions in only a part of the initial conditions, by creating a subgrid in a subregion, which is again convolved with the real space transfer function $T(\vec x_i)$. These are so-called {\em zoom initial conditions}. The description here is by no means sufficient to generate accurate zoom initial conditions, but is intended as a rough sketch. Refs.~\cite{Bertschinger:2001ng,Hahn:2011uy} discuss the matter in great detail, solving issues such as mass conservation and keeping the zoom region `invisible' to the coarsely sampled outer regions when it comes to the gravitational Poisson equation.

\subsection{Pre-initial conditions, or particles}
The choice of largest wave number $k=\pi M/L$ is unrelated to the choice of the number of particles, although it is tempting (and simplest) to set them equal, such that the vertices on the grid for the Fourier transform are directly interpreted as particles. This simplest choice is what we used in the illustrations in the body of this paper. An obvious deviation is to choose less particles, still on the vertices though, simply skipping a fraction of the generated vertices. 

The most popular choice is to put particles on {\em glass} pre-initial conditions. Particles are distributed randomly and evolved under a repulsive force with a friction to aid convergence, until they reach an equilibrium. Such a configuration has no preferred directions, as opposed to a grid. The downside is that the particles are not at the vertices of the Fourier grid, such that the values of the field at the particle positions must be obtained by interpolations, and a mathematically rigorous construction of the field values at particle positions is lost. The interpolation between the grid vertices, boils down to turning the discrete field back into a continuous one, by convolving the discrete field with a mass window function, the shape of which needs to be taken into account as it affects the power spectrum just like a filter. Other tilings, apart from glass and grid, exist as well~\cite{Hansen:2006hj}.

As pointed out in for example~\cite{Baertschiger:2001eu}, both grids and glasses are gravitationally unstable against small perturbations. This implies that part of the finest structure found in a simulation is due to the choice of pre-initial conditions, and not cosmological of origin.

\subsection{Summary}
The parameters that a simulator has to choose in the initial conditions, include at least:
\begin{itemize}
\item the box size $L$,
\item the highest wavenumber $\pi M/L$
\item additional filters on the initial spectrum (apart from the tophat at $\pi M/L$),
\item the number of vertices / particles, 
\item the position of the vertices (on the Fourier grid, glass, \ldots).
\end{itemize}
Each of these has its own consequences for the resolution of the simulation, and the meaning of its outcome. What these consequences are, is a whole field of research~\cite{Pen:1997up,Baertschiger:2001eu,Joyce:2004qz,Sirko:2005uz,Prunet:2008fv,Joyce:2008kg,Carron:2014wea,Colombi:2014zga}, and is beyond the scope of this paper.

}

\bibliographystyle{JHEP}
\bibliography{refs}
\end{document}